\documentclass{egpubl}
\usepackage{eg2025}

\usepackage[T1]{fontenc}
\usepackage{dfadobe}  

\usepackage{cite}  
\usepackage[dvipsnames]{xcolor}
\BibtexOrBiblatex
\electronicVersion
\PrintedOrElectronic
\ifpdf \usepackage[pdftex]{graphicx} \pdfcompresslevel=9
\else \usepackage[dvips]{graphicx} \fi

\usepackage{egweblnk} 

\title[Versatile Physics-based Character Control with Hybrid Latent Representation]%
      {Versatile Physics-based Character Control with Hybrid Latent Representation}

\author[J. Bae, J. Won, D. Lim, I. Hwang, Y.\,M. Kim]
{\parbox{\textwidth}{\centering
    Jinseok Bae\orcid{0000-0003-2409-290X},
    Jungdam Won\orcid{0000-0001-5510-6425},
    Donggeun Lim\orcid{0009-0000-1450-8428},
    Inwoo Hwang\orcid{0009-0005-9819-1873},
Young Min Kim\thanks{Corresponding Author}\orcid{0000-0002-6735-8539}
}
        \\
{\parbox{\textwidth}{\centering Seoul National University, South Korea
       }
}
}

\usepackage{algorithm2e}
\usepackage{amsmath}
\usepackage{amsfonts}
\usepackage{booktabs}
\begin{document}

\teaser{
 \includegraphics[clip=true, width=\linewidth]{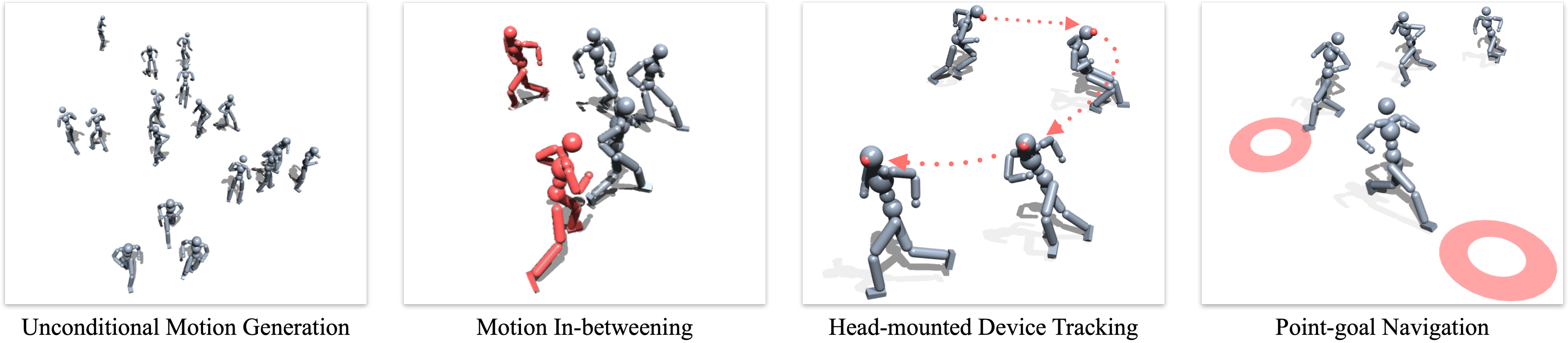}
 \centering
  \caption{
  Our hybrid latent representation serves as a versatile motion prior such that the agent can robustly adapt to a range of challenging downstream tasks.
  }
\label{teaserfigure}
}

\maketitle
\begin{abstract}
We present a versatile latent representation that enables physically simulated character to efficiently utilize motion priors.
To build a powerful motion embedding that is shared across multiple tasks, the physics controller should employ rich latent space that is easily explored and capable of generating high-quality motion.
We propose integrating continuous and discrete latent representations to build a versatile motion prior that can be adapted to a wide range of challenging control tasks.
Specifically, we build a discrete latent model to capture distinctive posterior distribution without collapse, and simultaneously augment the sampled vector with the continuous residuals to generate high-quality, smooth motion without jittering. 
We further incorporate Residual Vector Quantization, which not only maximizes the capacity of the discrete motion prior, but also efficiently abstracts the action space during the task learning phase.
We demonstrate that our agent can produce diverse yet smooth motions simply by traversing the learned motion prior through unconditional motion generation.
Furthermore, our model robustly satisfies sparse goal conditions with highly expressive natural motions, including head-mounted device tracking and motion in-betweening at irregular intervals, which could not be achieved with existing latent representations.


\end{abstract}
\section{Introduction}
Recent advances in physics-based character control have shown remarkable progress in generating realistic human motions using simulated agents and reinforcement learning~\cite{yuan2023learning, bae2023pmp, hassan2023synthesizing}.
Hierarchical control strategies~\cite{merel2018neural, merel2020catch, kumar2021rma} offer a scalable approach to obtain and utilize a general motion prior, instead of manually selecting reference data tailored for each scenario.
A hierarchical controller is trained in two phases: imitation learning and task learning.
In the imitation learning phase, an agent constructs a latent embedding to serve as a scalable motion prior. 
During the task learning phase, the agent explores this latent space to discover optimal skills for the given scenario.
Constructing a rich and structured latent space during the first phase is critical to enhancing training efficiency and motion quality during the task learning phase. 

A standard method extracts the latent space of motion through the bottleneck layer of an autoencoder.
For generative tasks, we often employ variational inference~\cite{won2022physics, yao2022controlvae, luo2023universal}, such that the latent representation matches a pre-defined probabilistic distribution, which is usually Gaussian.
However, variational frameworks frequently suffer from posterior collapse~\cite{lucas2019understanding}, failing to capture details from the original dataset.
Moreover, when combined with the hierarchical control, searching a latent vector within the continuous embedding space may be biased towards fulfilling the task reward and results in awkward motions that deviate from the original dataset.
As a remedy, several works build the latent space to be a structured set of discrete vectors~\cite{zhu2023neural, yao2023moconvq, guo2023momask}.
However, action sequences selected from the discrete latent space are temporally discontinuous, leading to artifacts that can significantly degrade motion quality.
Furthermore, these models face a critical trade-off: while a larger codebook increases the imitation policy's capacity, it complicates the exploration during the task learning phase.

We address these issues by integrating continuous and discrete latent space, namely \textit{hybrid} latent representation.
Unlike the continuous latent model, our model uses a discrete policy as a high-level strategy, ensuring that the generated motion thoroughly captures the diverse motion distribution in the dataset.
At the same time, our agent consistently produces smooth motions, enhancing visual quality compared to the agent with discrete latent representation. 
We attribute this to our novel policy architecture, which augments discrete motion prior with the temporally coherent motion prior from the continuous latent space.

Additionally, we improve the effectiveness of the discrete latent representation using Residual Vector Quantization (RVQ)\cite{lee2022autoregressive, zeghidour2021soundstream}. 
Our experiments show that RVQ improves the latent model in both the imitation and task learning phases.
During the imitation learning phase, RVQ significantly enhances the capacity of the imitation policy compared to a standard vector quantization layer.
By employing the \textit{quantizer dropout} technique~\cite{zeghidour2021soundstream} of RVQ, our method allows an agent to efficiently reduce the training complexity of the task learning phase by constraining the action space. 
We additionally suggest a strategy to increase controllability in task learning by balancing the exploration and the exploitation of learned motion prior.
While some works use RVQ for motion generation~\cite{guo2023momask, yao2023moconvq}, their approaches are not suited for challenging control tasks like motion tracking with extremely sparse targets, \textit{e.g.} head-mounted device tracking.
Their models encode sequences of kinematic states over a temporal window, which limits precise control in highly dynamic scenarios where the agent must respond quickly to the goal. 
In contrast, we use RVQ to learn discrete embeddings of short-term motion dynamics in neighboring timesteps. 
The enhanced temporal precision enables our latent model to create a control policy that exhibits robust responsiveness to environmental changes.

In our experiments, we demonstrate the versatility of our model across various downstream tasks. 
Our agent consistently produces high-quality motion while discovering optimal motion combinations from the latent space. 
The model outperforms existing approaches in challenging scenarios with sparse goals, such as motion in-betweening over arbitrary time intervals. 
Furthermore, our agent generates highly natural motion without explicit reinforcement for motion style. 
We also show that our novel latent representation enables the agent to easily solve tasks with simple reward designs, as seen in examples like head-mounted device tracking and point-goal navigation.

We summarize our contributions below.
\begin{itemize}
\item \textit{Improved stability and quality}. Our hybrid representation offers integrated latent space that significantly enhances training stability and motion quality for downstream tasks. 
\item \textit{Enhanced code usage and training}.
We utilize Residual Vector Quantization (RVQ) to maximize the usage of a limited number of discrete latent codes while reducing the training complexity during the task learning phase.
\item \textit{Versatility}.
We showcase our agent's capabilities across various downstream tasks, demonstrating that the proposed latent embedding can discover optimal motion combinations in challenging unseen target scenarios.
\end{itemize}

\section{Related Works}
\subsection{Character Animation}
Advancements in the data-driven approach demonstrated that virtual characters can learn natural, human-like motion from large datasets.
With a sufficient number of samples from extensive motion datasets~\cite{ionescu2013human3, mahmood2019amass, harvey2020robust, lin2024motion}, neural networks can understand patterns in human motion.
Recent studies have demonstrated the ability to reconstruct complex interactions involving humans and objects~\cite{peng2023hoi, diller2023cghoi, xu2024interdreamer, zhang2024hoi, bhatnagar22behave}, humans and other humans~\cite{liang2024intergen, shafir2023human, xu2024regennet}, and humans within environments~\cite{jiang2024scaling, wang2022humanise, Zhao:ICCV:2023, yi2024tesmo}.

On the other hand, physics-based control generates a sequence of actions relying on physics simulators to produce physically plausible, therefore highly realistic motion.
Pioneering works generated high-quality motion of virtual characters by training imitation policies using deep reinforcement learning (DRL)~\cite{peng2018deepmimic, xu2021gan, peng2021amp}. 
The approaches are further extended to generate complex physical interactions between agents~\cite{younes2023maaip} or between agents and objects~\cite{bae2023pmp, wang2023physhoi, hassan2023synthesizing}.
However, training DRL controllers from scratch for each scenario is not sample-efficient, making the process computationally demanding.

A stream of research suggests enhancing the efficiency of controllers by employing reusable skill prior that is built from motion datasets~\cite{merel2018neural, merel2020catch, kumar2021rma, won2021control}.
These methods often use a hierarchical strategy. 
First, a low-level controller is trained, which extracts a latent space that encapsulates reference motions.
Then, the high-level policy explores the latent space and outputs the latent vectors, which are then decoded into physical actions by the pre-trained low-level controller.
A line of works have employed continuous latent space~\cite{ling2020character, won2022physics, yuan2023learning, luo2023universal, peng2022ase, tessler2023calm, yao2022controlvae, xu2023adaptnet}, while some recent works~\cite{zhu2023neural, yao2023moconvq} have leveraged discrete latent representation.
However, it is not trivial to obtain stable yet rich latent space which allows structured exploration to adapt to novel scenarios with natural motion.
We further examine the limitations of these approaches in Sec.~\ref{sec:03_preliminaries}

\subsection{Latent Representation}
Variational Autoencoders (VAE)~\cite{kingma2013auto} have been foundational for generative modeling, particularly in handling complex data distributions with continuous latent vectors. 
While VAEs have been successfully applied across various domains~\cite{chung2015recurrent, gregor2018temporal, tevet2022human}, they often face stability issues during training.
This instability largely stems from the need to balance the trade-off between reconstruction fidelity and the regularization imposed by KL divergence~\cite{bowman2015generating,higgins2017beta}.
As a partial remedy, one needs to adjust weights with proper scheduling, which requires complex tuning.

Building on the principles of VAE, Vector Quantized Variational Autoencoders (VQ-VAE)~\cite{van2017neural, razavi2019generating} introduce a significant evolution in the field of latent representations.
By converting continuous data into distinct vectors, VQ-VAEs stabilize training and enhance output clarity, making them ideal for tasks like video and audio synthesis~\cite{garbacea2019low, yan2021videogpt}.
Recent studies have further advanced network architecture for VQ-VAE, mainly focusing on increasing the code usage~\cite{zeghidour2021soundstream, yang2023hifi, yu2021vector, lee2022autoregressive} and the efficiency of the quantization process~\cite{mentzer2023finite, yu2023language}.
However, VQ-VAEs have limitations, such as overfitting to small datasets and challenges in increasing quantization levels without compromising model efficiency.
Additionally, when discrete codes from a VQ-VAE structure are used for hierarchical control \cite{zhu2023neural}, increasing the codebook size directly enlarges the action space, which undesirably raises the training complexity.
\begin{figure}[t]
    \centering
    \includegraphics[clip=true, width=0.90\linewidth]{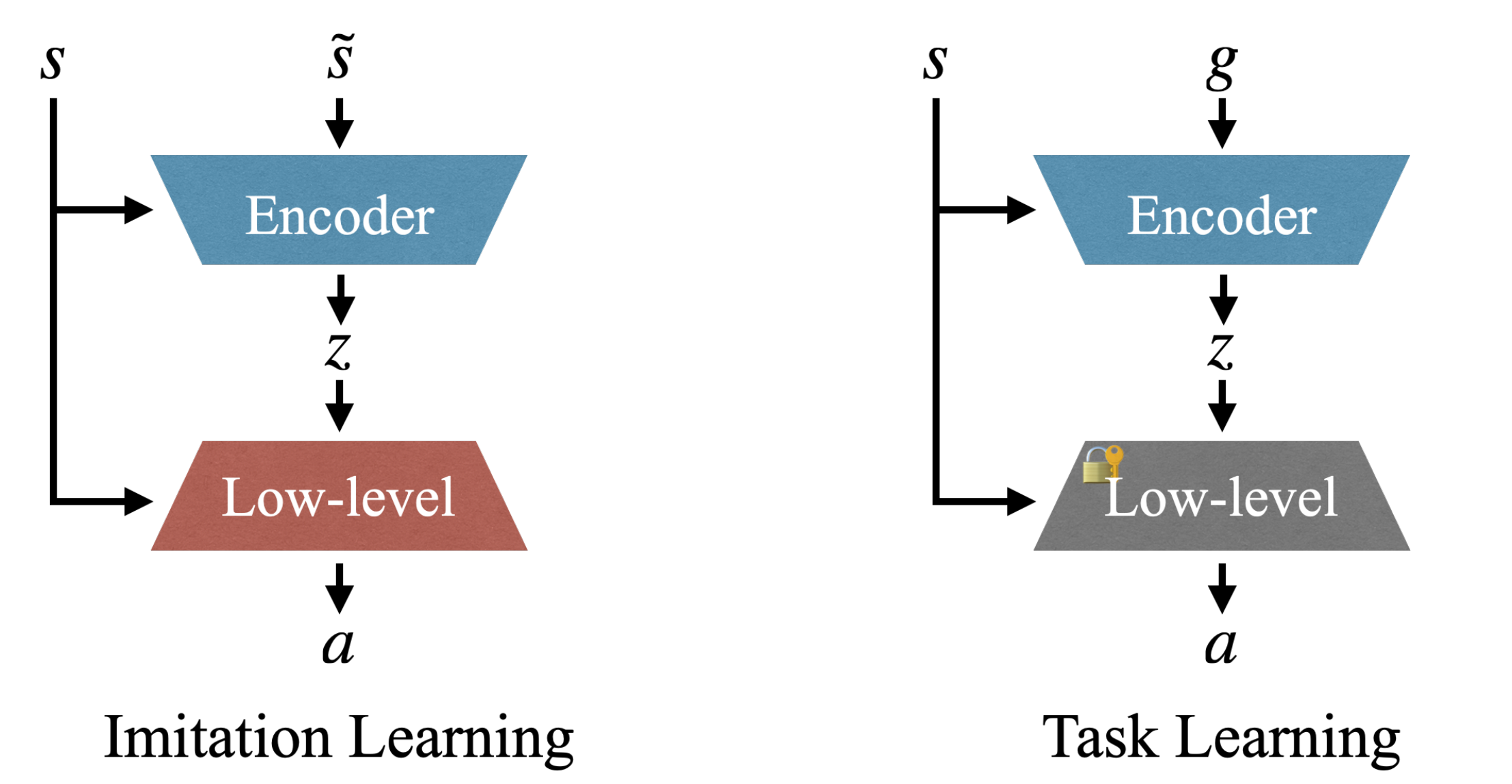}
    \caption{Overview of the hierarchical control. (\textit{Left}) At the first stage of imitation learning, an encoder and the low-level policy are jointly trained to learn latent space for a prior. (\textit{Right}) During the task learning phase, high-level policy is trained to output proper latent action for the pretrained low-level policy.}
    \label{fig:hier_control}
\end{figure}  

\section{Preliminaries}\label{sec:03_preliminaries}
In this section, we briefly explain the base framework for hierarchical control (Sec.~\ref{sec:03_01_hierarchical}).
Then, we review two types of motion priors, namely \textit{continuous} (Sec.~\ref{sec:03_02_continuous}) and \textit{discrete} (\ref{sec:03_03_discrete}) priors.

\subsection{Hierarchical Control}\label{sec:03_01_hierarchical}
Hierarchical control enables an agent to reuse pretrained motion priors in various types of downstream tasks.
The training pipeline can be split into two stages: the first stage to train the low-level controller $\pi_{\text{low}}$, and the second stage to train the high-level controller $\pi_{\text{high}}$.
In the first stage, $\pi_{\text{low}}$ is trained to output appropriate actions $a$ given a goal state $\tilde{s}$ sampled from reference motion clips.
Here, the policy learning employs an encoder-decoder structure such that the agent can learn a compact yet expressive representation $z$ for each pair of $(s, \tilde{s})$.
Specifically, the encoder maps the requested transition $(s,\tilde{s})$ to a latent embedding $z$, while $\pi_{\text{low}}$ decodes $s$ and $z$ to output action $a$.
Once trained, the parameters of the decoder are frozen.

At the second stage, the framework trains a new encoder, the high-level policy $\pi_{\text{high}}$, that outputs desirable latent vector $z$ given $s$ and the task-specific goal $g$.
Therefore, the pretrained latent space must faithfully represent the dataset, while allowing $\pi_{\text{high}}$ to safely explore the latent space.
We illustrate the overall process of the hierarchical control in Figure~\ref{fig:hier_control}.

\subsection{Continuous Prior}\label{sec:03_02_continuous}
Regularization technique can further improve a basic encoder-decoder structure~(Figure\ref{fig:hier_control}) with comprehensive latent space.
Previous works~\cite{won2022physics, yao2022controlvae, luo2023universal} suggest a regularization technique inspired by the VAE structure~\cite{kingma2013auto}.
Specifically, these methods allow the encoder to output a posterior distribution $q_\phi$, and sample a latent vector $z$ from $q_\phi$.
Conventionally, the \textit{Kullback-Leibler} (KL) divergence loss $\mathcal{L}_{\text{KL}}$ regularizes $q_\phi$ to match a desired continuous distribution $p_\theta$, which is either a standard normal distribution $\mathcal{N}(0,I)$ or a distribution parameterized by an additional prior network $f_\theta$ that is conditioned to $s$.
We summarize the behavior of $\pi_{\text{low}}$ and $\mathcal{L}_{\text{KL}}$ in the first stage as
\begin{equation}\label{eq:continuous_low_level}
    \begin{gathered}
    a \sim \pi_{\text{low}}(a|s, z),~\text{where}~z \sim q_\phi(z|s,\tilde{s}),\\
    \mathcal{L}_{\text{KL}} = \text{KL}(q_\phi(z|s,\tilde{s})|| p_\theta(z|s)).
    \end{gathered}
\end{equation}
During the high-level policy training, a previous work~\cite{luo2023universal} empirically suggests outputting a residual value  $u$ relative to the mean of the prior distribution $\mu_p=f_\theta(s)$ produces better results than directly outputting $z$ on downstream tasks.
Then the latent action $z$ from $\pi_{\text{high}}$ in the second stage can be written as
\begin{equation}
    \label{eq:continuous_high_level}
    z = \mu_p + u,~\text{where}~u\sim\pi_{\text{high}}(u|s, g).
\end{equation}

We denote this type of method as a \textit{continuous} prior.
Although the \textit{continuous} prior allows the agent to generate smooth trajectories, the agent may experience posterior collapse, which makes the latent representation $z$ less informative.
Furthermore, the model requires extensive hyperparameter tuning, such as noise level for proper exploration, to stabilize the high-level policy training. 
Otherwise, we find that the agent with \textit{continuous} prior easily generates unnatural motion that deviates from the valid motion space in the embedded prior, especially when the reward for the downstream task does not consider motion quality (\textit{e.g.}, \textit{point-goal navigation} that employs a reward for the final position only).

\subsection{Discrete Prior}\label{sec:03_03_discrete}
Previous works~\cite{zhu2023neural, yao2023moconvq} suggest the vector quantization approach~\cite{van2017neural} to overcome drawbacks of \textit{continuous} motion prior.
We categorize these models as \textit{discrete} models and explain the behavior of their motion prior based on Neural Categorical Prior (NCP)~\cite{zhu2023neural}.
The posterior network $f_\phi$ initially outputs a continuous latent vector $z$, and then quantizes it as $\bar{z}$ by selecting the nearest code $e^*$ from the codebook $\mathcal{B}$.
This can be viewed as a mapping process that forces the quantizer to commit to a specific vector in the codebook for each input.
The commitment loss $\mathcal{L}_{\text{commit}}$ in the VQ-VAE structure~\cite{van2017neural} attaches the latent vectors to a code while updating the codebook parameters:
\begin{equation}\label{eq:vector_quantization}
    \begin{gathered}
  e^* = \underset{e\sim \mathcal{B}}{\mathrm{argmin}}{\|e-z\|^2},~\text{where}~z = f_\phi(s, \tilde{s}),\\
  \mathcal{L}_{\text{commit}} = \|z - \text{sg}(e^*)\|^2 + \|\text{sg}(z) - e^*\|^2,
  \end{gathered}
\end{equation}
where $\text{sg}(\cdot)$ denotes the stop-gradient function.
Note that vector quantization assumes a uniform distribution as a prior distribution, so the KL divergence term becomes constant.
Therefore, the training objective can alleviate KL divergence loss, which significantly accelerates the process without time-consuming hyperparameter tuning.
Finally, the \textit{straight-through} trick~\cite{van2017neural} enables back-propagation through the quantized latent $\bar{z}$, which can be written as
\begin{equation}\label{eq:straight-through}
  \bar{z} = z + \text{sg}(e^* - z).
\end{equation}
Given $\bar{z}$ and $s$, the low-level controller $\pi_{\text{low}}$ predicts the action.

At the task learning phase, the \textit{discrete} model treats the high-level policy $\pi_{\text{high}}$ as a discrete policy, which chooses the index of a desirable embedding $e$ from the pretrained codebook $\mathcal{B}$.
Since $\pi_{\text{high}}$ explores a finite number of discrete actions, the model prevents $\pi_{\text{high}}$ from deviating largely from the pretrained motion prior.
However, the discrete choice of the latent code triggers high-frequency jittering in motion since selected codes can be temporally discontinuous.
Similarly, the motion generated from random prior sampling often encounters jerky movements.
To employ a better prior distribution, \textit{discrete} latent model may allocate an additional training stage for the prior network and use it as a regularizer for downstream tasks.
However, we find that the prior network cannot imitate the dataset as much as the posterior network, especially when the motion dataset is large.
The reported training curve from NCP~\cite{zhu2023neural} also supports that the performances in the tasks are nearly the same with or without additional prior network.
\begin{figure}[t]
  \centering
  \includegraphics[clip=true, width=0.93\linewidth]{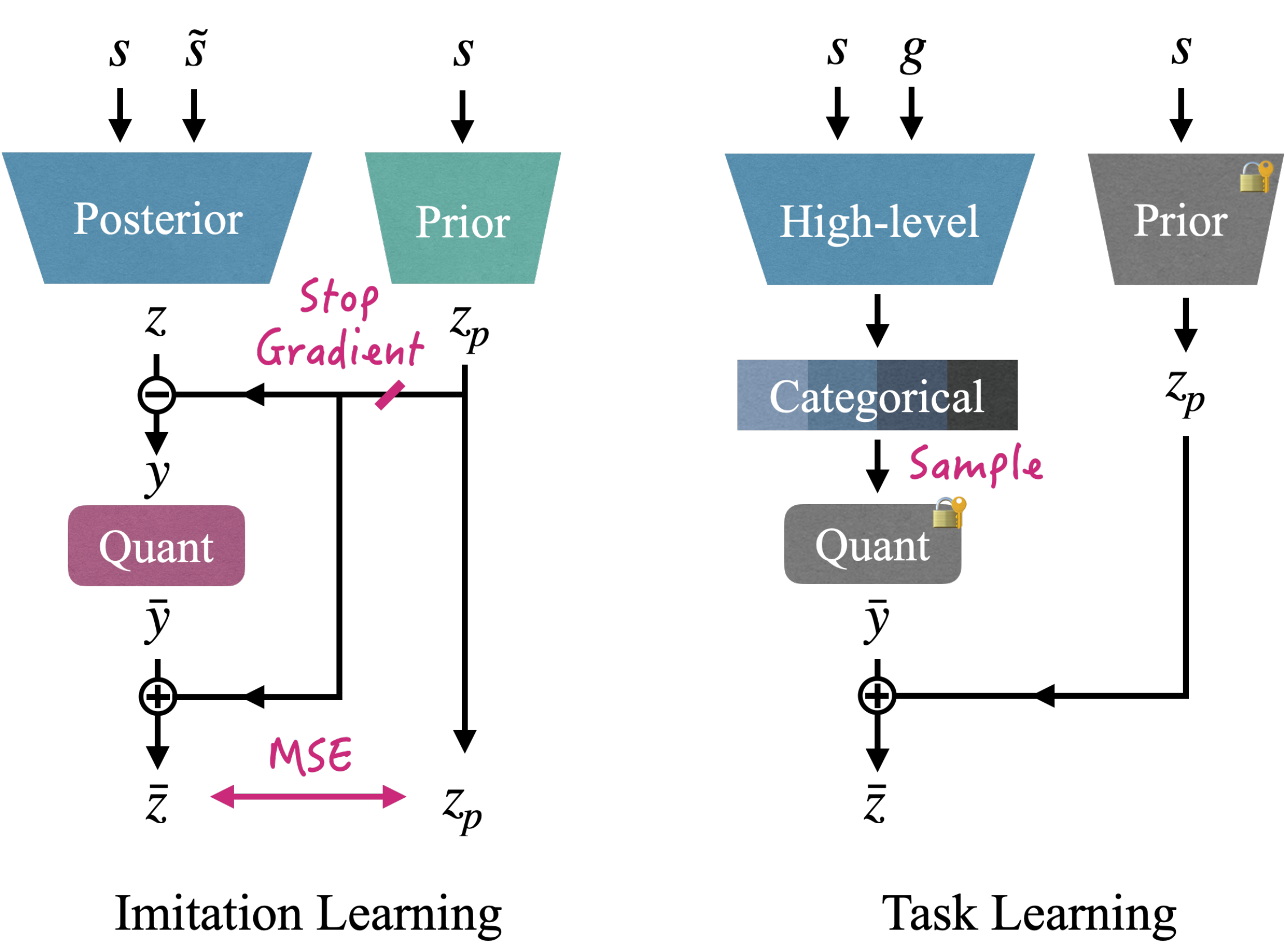}
  \caption{Network architecture for \textit{hybrid} motion prior. (\textit{Left}) During the imitation learning phase, posterior and prior network output latent vector $z$ and $z_p$ given the current state $s$ and target state $\tilde{s}$. Next, the network quantizes the marginal value of $y=z-z_p$ into $\bar{y}$, then reconstructs the final latent representation $\bar{z}$ by biasing the quantized value with  $z_p$. Once trained, the parameters of the codebook and prior network are frozen. (\textit{Right}) In the task learning phase, high-level policy outputs categorical distribution, and then samples the index for the pretrained codebook. The latent $\bar{z}$ is then reconstructed with the selected $\bar{y}$ and continuous vector $z_p$.}
  \label{fig:our_architecture}
\end{figure}  

\section{Method}
In this section, we detail our novel architecture, which contributes to smooth motion trajectories through efficient training across a variety of downstream tasks. 
We refer to the motion prior derived from our model as \textit{hybrid} motion prior, as it benefits from both \textit{continuous} and \textit{discrete} priors.
We first describe the \textit{hybrid} motion prior learned from the imitation phase (Sec.~\ref{sec:04_01_mq}), followed by a method to enhance \textit{discrete} prior using Residual Vector Quantization (Sec. ~\ref{sec:04_02_rvq}).
Lastly, we elaborate on the improved training procedure for the imitation learning and the task learning phases (Sec.~\ref{sec:04_03_training}).

\subsection{Hybrid Prior using Marginal Quantization (MQ)}\label{sec:04_01_mq}

As shown in Figure~\ref{fig:our_architecture}, the prior network outputs continuous latent vector $z_p$ to maintain the expressivity and temporal smoothness, while the posterior and high-level policy networks achieve stability by generating categorical outputs that are discretized.
We introduce Marginal Quantization (MQ), a method that quantizes only the difference between the posterior and prior latent vectors.
The process begins by computing the margin $y = z - \text{sg}(z_p)$, where $z$ and $z_p$ are the latent vectors from the posterior and prior networks, respectively. 
The margin $y$ is then quantized to $\bar{y}$, and the final latent vector produced is $\bar{z} = \bar{y} + \text{sg}(z_p)$. 
The attached quantization potentially avoids posterior collapses and encourages the posterior network to output informative continuous latent $z$ to achieve accurate imitation.
Nonetheless, training the final value of $\bar{z}$ observes the original gradient calculated for the continuous output $z$ as the subtracted value of $\text{sg}(z_p)$ is added back at the end.
This is an extension of the \textit{straight-through} technique, originally developed to enable back-propagation during vector quantization~\cite{van2017neural}.
Attaching the discrete vector $\bar{z}$ to the continuous vector $z_p$ from the prior network, the final vector $\bar{z}$ stays temporally coherent throughout an episode during the task learning phase.

Additionally, we incorporate a margin-minimizing loss $\mathcal{L}_{\text{mm}}$ to align the posterior with the prior latent space, defined as $\mathcal{L}_{\text{mm}} = \|y\|^2 = \|\bar{z} - z_p\|^2$.
Unlike $\mathcal{L}_{\text{KL}}$ in the \textit{continuous} model, we empirically find that the overall performance is not sensitive to the weight multiplied by the additional loss $\mathcal{L}_{\text{mm}}$.
We summarize the behavior of our imitation policy in Algorithm~\ref{alg:marginal_quantization}.

\RestyleAlgo{ruled}
\SetKwComment{Comment}{$\rhd$ }{}
\SetKwInput{KwNetwork}{Network}
\SetKwInput{KwDef}{Definition}
\SetKwFunction{FMain}{ForwardActor}
\SetKwProg{Fn}{Function}{:}{\KwRet $a,~\mathcal{L}_{\text{commit}},~\mathcal{L}_{\text{mm}}$}
\DontPrintSemicolon
\begin{algorithm}
\caption{Action Prediction using MQ}\label{alg:marginal_quantization}
\KwNetwork{posterior network $f_\phi$, prior network $f_\theta$, quantizer $Q$, low-level controller $\pi_{\text{low}}$}
\KwDef{stop gradient $sg(\cdot)$, latent $z$, margin $y$}
\KwIn{current state $s$, target reference state $\tilde{s}$}
\KwOut{action $a$, commitment loss $\mathcal{L}_{\text{commit}}$, margin-minimizing loss $\mathcal{L}_{\text{mm}}$}
\Fn{\FMain{$s$, $\tilde{s}$}}{
    $z\leftarrow f_\phi(s, \tilde{s})$ \Comment*[r]{\small latent vector from posterior}
    $z_p\leftarrow f_\theta(s)$ \Comment*[r]{\small latent vector from prior}
    $y\leftarrow z - sg(z_p)$ \Comment*[r]{\small calculate margin}
    $\bar{y},~\mathcal{L}_{\text{commit}}\leftarrow Q(y)$ \Comment*[r]{\small calculate $\bar{y},~\mathcal{L}_{\text{commit}}$}
    $\bar{z}\leftarrow \bar{y} + sg(z_p)$ \Comment*[r]{\small reconstruct $z$}
    $\mathcal{L}_{\text{mm}}\leftarrow \|\bar{z} - z_p\|^2$ \Comment*[r]{\small calculate $\mathcal{L}_{\text{mm}}$}
    $a\sim \pi_{\text{low}}(s, \bar{z})$ \Comment*[r]{\small sample $a$}

}
\end{algorithm}

\subsection{Enhanced Code Usage for Discrete Prior}\label{sec:04_02_rvq}
As we develop versatile motion prior that can adapt to various challenging tasks, it is crucial to allow efficient exploration within the rich set of plausible motion space.
Especially the discrete part of our hybrid representation leverages VQ-VAE~\cite{van2017neural,zhu2023neural}, whose capacity, defined by the size of the codebook $\mathcal{B}$, is not fully utilized due to the low rate of code usage~\cite{mentzer2023finite}.
To address this, we employ hierarchical codebooks of Residual Vector Quantization (RVQ)~\cite{lee2022autoregressive,zeghidour2021soundstream}, which sequentially uses $N$ codebooks and retrieves discrete latent codes in an auto-regressive manner.
Our residual quantization process $Q$ for margin $y$ is represented as
\begin{equation}\label{eq:residual_quantization}
  \bar{y} = \sum_{n=1}^N{e^*_n},~\text{where}~e^*_n=\underset{e_n\sim \mathcal{B}_n}{\mathrm{argmin}}{\|e_n-(y - \sum_{k=1}^{n-1}{e^*_k})\|^2}.
\end{equation}
We find that even with the same total number of codes, a model with RVQ provides better imitation quality.
In addition, we utilize three training techniques for RVQ used in the previous works~\cite{zeghidour2021soundstream, ao2022rhythmic, yao2023moconvq, jiang2024motiongpt}: \textit{Exponential Moving Average (EMA) update} for optimizing the codes, \textit{code reset} technique for eliminating unused codes, and the \textit{quantizer dropout} to build priority on the multiple codebook system.
Among the techniques, \textit{quantizer dropout} creates a critical advantage for the task learning phase as further explained in Section~\ref{sec:04_03_training}.

\subsection{Efficient Training for Policies}\label{sec:04_03_training}
\subsubsection{Imitation Learning}
Our training pipeline begins with the imitation learning phase where an agent learns latent motion prior via imitating the motion capture data.
We find that conventional RL is less efficient at training complex latent models compared to simpler models, as shown in Figure~\ref{fig:hier_control}.
This is because latent models require additional terms, such as KL divergence or commitment losses, to regularize the latent space.
To address this, we adopt \textit{online distillation}\cite{luo2023universal} as a simple yet effective approach for training a low-level policy.
In online distillation, a student policy is trained under the supervision from a pretrained expert policy. 
The online distillation enables the student policy to achieve highly accurate imitation, which would not be possible if trained from scratch with RL.
Following the previous work~\cite{luo2023universal}, we first train an expert policy using Proximal Policy Optimization (PPO) algorithm.
As an expert policy, we use a simple encoder-decoder structure as illustrated in Figure~\ref{fig:hier_control}.
We explain further details about expert policy in the Appendix.
Then we train an agent with supervised guidance from the pretrained expert imitation policy.
We add the commitment loss and the margin-minimizing loss to the original training objective~\cite{luo2023universal} as
\begin{equation}\label{eq:imitation_objective}
  \mathcal{L}=\mathcal{L}_{\text{action}} + \alpha\mathcal{L}_{\text{reg}} + \beta\mathcal{L}_{\text{commit}} + \gamma\mathcal{L}_{\text{mm}},
\end{equation}
where $\mathcal{L}_{\text{action}}$ refers to supervised guidance from expert action $\|a-a_\text{expert}\|^2$, while $\mathcal{L}_{\text{reg}}$ is a regularization term to minimize change between latent embeddings of neighboring frames. 
We additionally modify the regularization term as $\mathcal{L}_{\text{reg}}=\|\bar{y} - \bar{y}'\|^2 + \|z_p - z'_p\|^2$ where $\bar{y}'$ is the quantized margin and $z'_p$ is the prior latent vector cached from the previous time step.

Since our model employs RVQ, the latent representation better utilizes the limited number of codes compared to the conventional \textit{discrete} model (Sec.~\ref{sec:04_02_rvq}). 
However, as a trade-off, RVQ exponentially increases training complexity during the task learning phase as our $\pi_{\text{high}}$ needs to explore the combinatorial action space composed of multiple codebooks.
To mitigate this problem, we utilize \textit{quantizer dropout} technique~\cite{zeghidour2021soundstream, yao2023moconvq} while training the codebooks.
Specifically, we randomly choose the maximum codebook index $M\in[1, N]$ where $N$ represents the total number of codebooks.
We then only use $\{\mathcal{B}_1, \cdots, \mathcal{B}_M\}$ to reconstruct the target value, which is equivalent to Eq.~(\ref{eq:residual_quantization}), but with $N$ in the original equation substituted by $M$.
As a result, our agent prioritizes codebooks with lower indices, thereby establishing a hierarchical structure within the overall set of codebooks.

\subsubsection{Task Learning}
In the task learning phase, our agent efficiently navigates the latent space for motion prior and robustly adapts to each task scenario with a high-level policy $\pi_{\text{high}}$.
The prioritization among codebooks allows $\pi_{\text{high}}$ to reduce the action space efficiently.
$\pi_{\text{high}}$ can even perform only with the highest priority codebook, $\mathcal{B}_1$, in downstream tasks.
This significantly improves over conventional discrete models, which must investigate all available codes to optimize the action space.
Additionally, RVQ provides extra controllability during task learning. Empirically, we find that using more codebooks for $\pi_{\text{high}}$ allows the agent to exhibit more diverse motions, encouraging exploration during training.
However, using too many codebooks can increase training complexity, so users can select the proper number to balance the results.
\section{Experiments}
We design several experiments to showcase the extreme versatility of our model with enhanced motion quality and training efficiency.
We additionally provide the qualitative results with the supplementary video and further ablation studies in the Appendix.

\subsection{Implementation Details}
We train our agent in parallel environments on Isaac Gym simulator~\cite{makoviychuk2021isaac}, with the internally implemented proportional derivative (PD) controller.
In each time step, the environment calculates state vector $s$, which can be written as
\begin{equation}
    s = \{h, q_r,\dot{p}_r,\dot{q}_r, q, \dot{q}, p_{\text{ee}}\},
\end{equation}
where $h, q_r,\dot{p}_r,\dot{q}_r$ denote height, orientation, velocity, and angular velocity of the root in character coordinate, respectively. 
Also, $q, \dot{q}$ represents the position and velocity of joints, while $p_{\text{ee}}$ refers to the end-effector position in character coordinates.
Also, we allow low-level policy $\pi_{\text{low}}$ to output action $a$ that acts as target joint positions $\hat{q}$ for PD control.
Following the previous work~\cite{zhu2023neural}, we design $\pi_{\text{low}}$ to output the residual action relative to the current state of joint position $q$, which can be represented as $\hat{q}=q+a$.
We train the models with LaFAN1 dataset~\cite{harvey2020robust} for imitation learning, which contains 3.5 hours of diverse motion capture data.

\subsection{Baselines}
We compare our \textit{hybrid} latent space model with representative works that use latent space of motion prior.
We adopt PULSE~\cite{luo2023perpetual} and NCP~\cite{zhu2023neural} from the released codes, and use them as a \textit{continuous} and \textit{discrete} models, respectively. 
We verify that our implementations perform best using most of the hyperparameters reported in the original papers.
This indicates that our implementation faithfully reproduces the original versions of the previous works. 
Nonetheless, we carefully tune some hyperparameters for the baselines to achieve the best quality for imitation and task learning. 
Additionally, we find the \textit{discrete} model can be further improved by incorporating Residual Vector Quantization (RVQ) layers. 
Therefore, we also implement the \textit{discrete} latent model enhanced with RVQ, which we denote as \textit{discrete$^+$} model.
We assign a codebook with the 8192 codes for the \textit{discrete} model, while the RVQ models (both \textit{discrete$^+$} and \textit{hybrid}) use the same total number of codes but in $N=8$ codebooks with 1024 codes each.
During the task learning, we manually tune the optimal hyperparameters for the exploration of $\pi_{\text{high}}$ in \textit{continuous} model.

\section{Discussions}
\subsection{Motion Imitation}
\begin{figure}[t]
    \centering
    \includegraphics[clip=true, width=\linewidth]{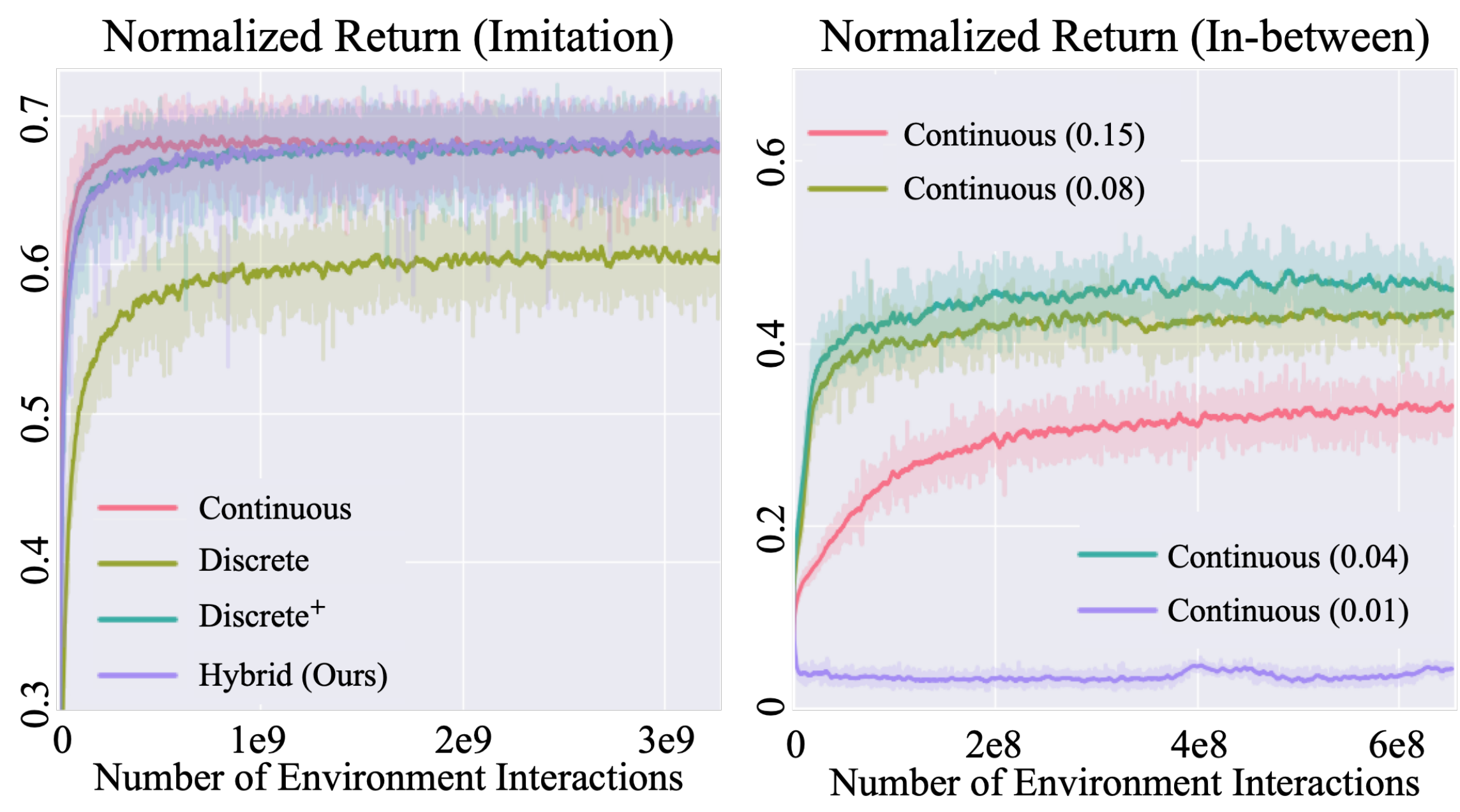}
    \caption{Training curves during the imitation and the task learning phases, respectively. (Left) \textit{Discrete} model with simple vector quantization cannot achieve accurate imitations. In contrast, \textit{discrete$^+$} model exhibits a similar training curve to ours. (Right) The value inside the parenthesis indicates experimented hyperparameters for the exploration rate. Continuous latent models are susceptible to the hyperparameters.}\label{fig:training_curve}
\end{figure}
We first examine the capacity of the various latent representations by testing the high-level policy to imitate the reference motions in the dataset.
Figure~\ref{fig:training_curve} shows the quality of imitation in terms of the reward function.
As suggested by previous works~\cite{luo2023universal}, continuous latent space can closely capture a diverse set of motions. 
In contrast, a finite set of discrete vectors often cannot comprehensively cover the dataset despite increased stability in various downstream tasks.
Although increasing the number of codes can slightly improve imitation quality, we cannot achieve a comparable level of imitation with the \textit{discrete} model.
We find that using RVQ can overcome the prevalent limitation of \textit{discrete} model, significantly improving the imitation quality.
Despite using the same number of codes with \textit{discrete }model, RVQ (\textit{discrete}$^+$, \textit{hybrid}) achieves the imitation quality on par with \textit{continuous} latent space by efficiently facilitating the finite number of codes.
Similar to \textit{discrete} latent model, we find the imitation quality to be less accurate when we use the hybrid latent model with the standard VQ layer. We further provide a comparison between the hybrid latent model using RVQ and the standard VQ in the Appendix.
\subsection{Unconditional Motion Generation}\label{sec:06_02_unconditional}
\begin{figure}[t]
    \centering
    \includegraphics[clip=true, width=0.95\linewidth]{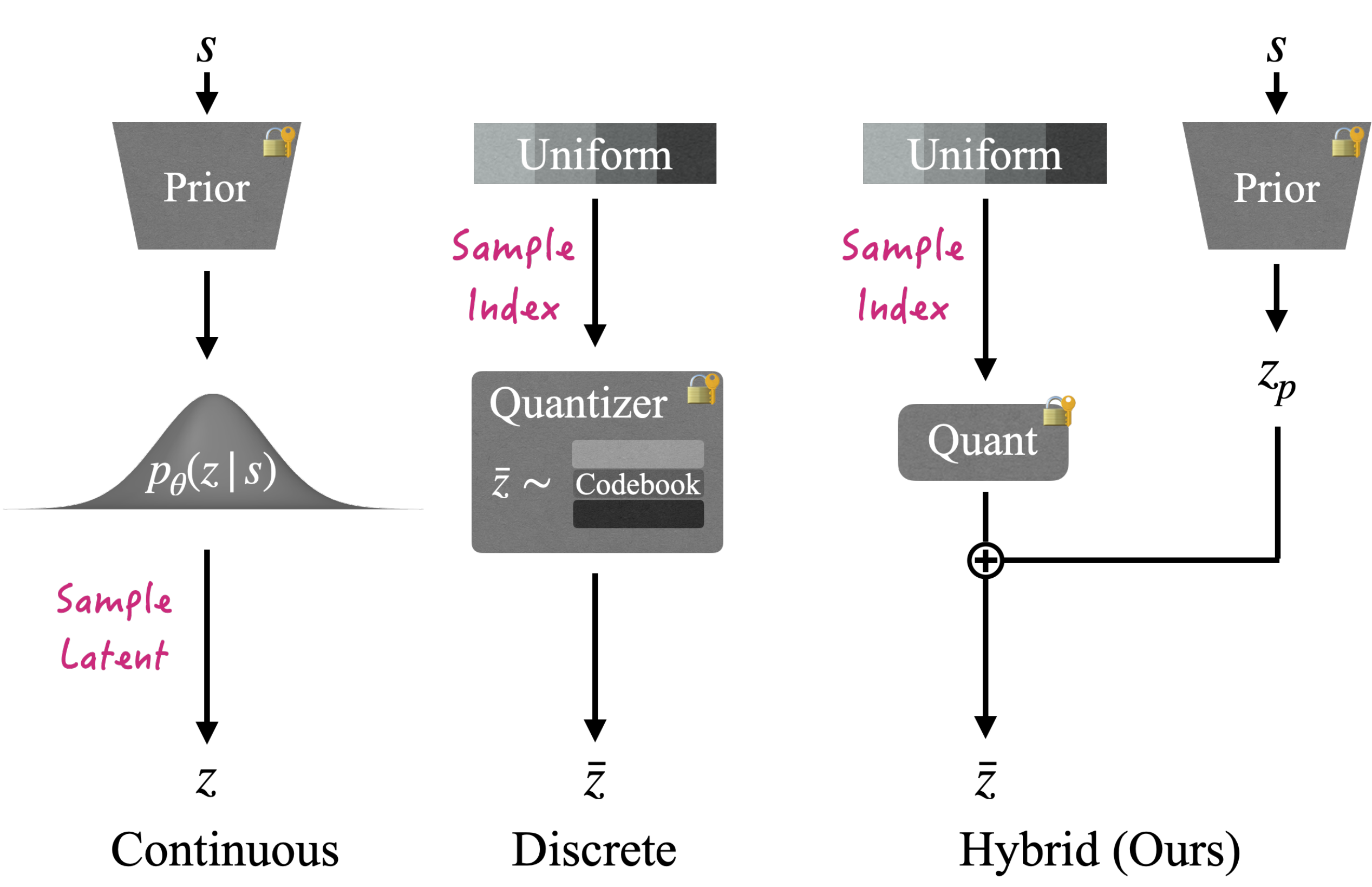}
    \caption{Prior sampling strategies for three different architectures. While \textit{continuous} model samples are latent directly from the parameterized Gaussian distribution $p_\theta$, \textit{discrete} and \textit{hybrid} models uniformly sample the index from the pretrained codebook.}\label{fig:prior_sampling}
\end{figure}
\begin{figure}[t]
    \centering
    \includegraphics[clip=true, width=\linewidth]{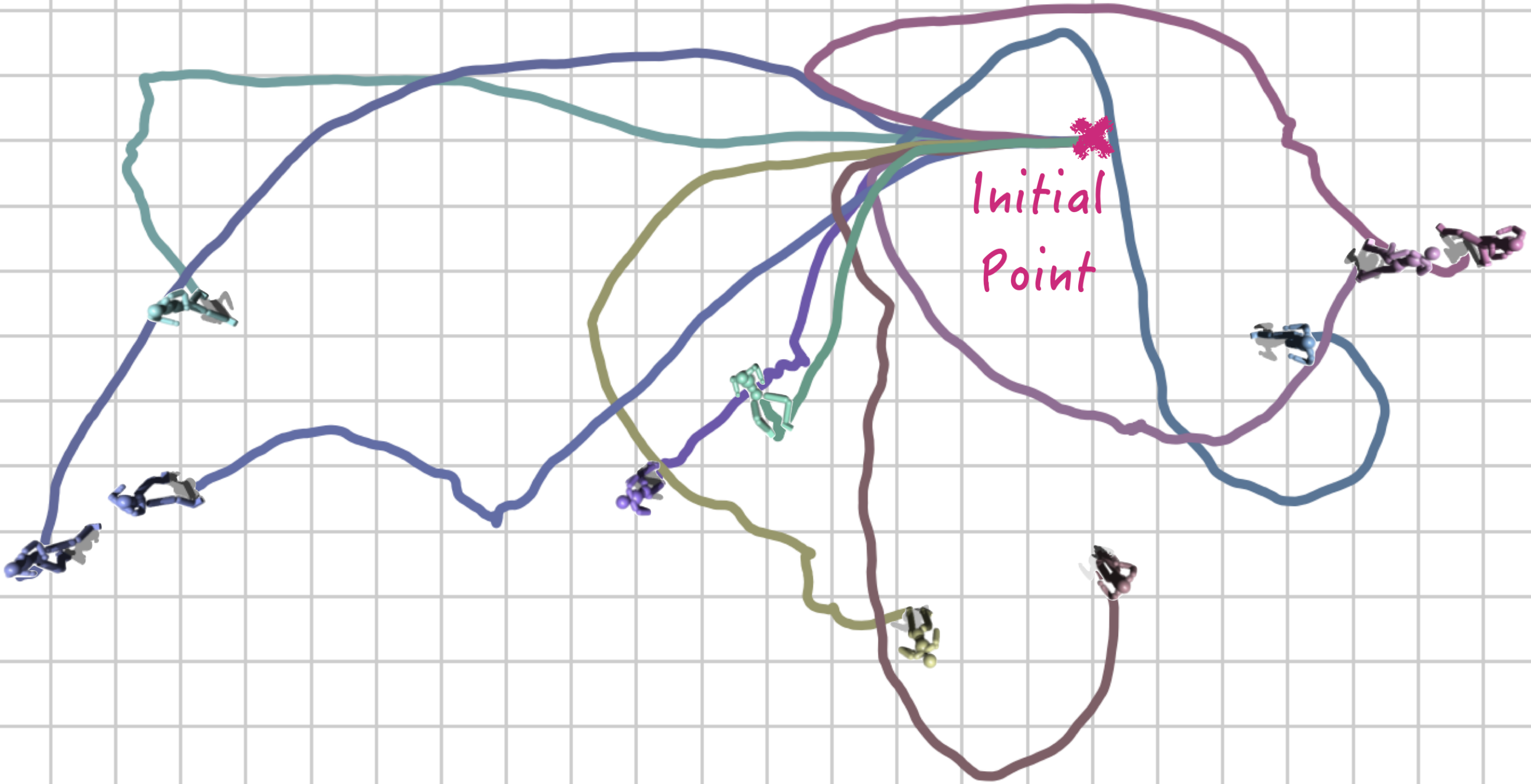}
    \caption{Unconditional motion generation from our model. We visualize trajectories from 30 seconds of 10 random rollouts with identical initial states.}
    \label{fig:prior_rollout_viz}
\end{figure}

We additionally evaluate the richness of the learned latent space by unconditional motion generation.
In this setup, we randomly sample the latent vector $z$ from the prior distribution of each model, as depicted in Figure~\ref{fig:prior_sampling}.
The prior distribution of the \textit{continuous} model is Gaussian distribution parameterized by its prior network.
In contrast, the \textit{discrete} and \textit{hybrid} models leverage VQ-VAE~\cite{van2017neural}, which utilizes a uniform distribution as its prior distribution.
We visualize the generated trajectories from 10 agents started with the same initial states in Figure~\ref{fig:prior_rollout_viz}.
Surprisingly, our agents produce extremely smooth motions throughout the rollouts, even though the random sampling for the latent vector is completely independent at each timestep.
We attribute this to our continuous latent vector $z_p$ from the prior network, which contributes to the temporal coherence.
Due to the absence of such a continuous latent vector, we find that the \textit{discrete} models largely suffer from high-frequency jittering.

We provide quantitative evaluations for the generated motions using the motion matching approach~\cite{peng2022ase} in three metrics. 
The \textit{distance} $d$ measures the similarity between the generated and reference motions regarding the Euclidean distance between state transitions.
For every pair of $(s, s')$, where $s$ and $s'$ are the current and the past proprioceptive states, we find the distance to the nearest state transitions $(s_m, s'_m)$ in the dataset $\mathcal{M}$
\begin{equation}
    d(s,s') = \min_{(s_m, s'_m) \in \mathcal{M}}{\|s-s_m\|^2 + \|s'-s'_m\|^2}.
\end{equation}
Following the previous work~\cite{peng2022ase}, we normalize each dimension of the state vector with the mean and standard deviation calculated from the dataset.
We also set a threshold value; if $d$ exceeds this threshold, we regard the queried transition as invalid and filter the calculated distance from the final statistics.
We also report the \textit{filtered rate} to show the ratio of valid motion in generations.
Finally, the \textit{coverage} counts the total number of visited transitions among datasets, representing the generated motion's diversity.

\begin{small}
\begin{table}
    \centering
  \caption{Evaluation on the unconditional motion generation. For \textit{discrete$^+$} and \textit{hybrid} model with RVQ, we indicate number of the used codebook inside parenthesis. We additionally mark the second best value with underlines.}
  \label{tab:prior_rollout}
  \resizebox{0.90\linewidth}{!}{
  \begin{tabular}{c|ccc}
    \toprule
    Models & Distance ($d$) $\downarrow$ & Filtered (\%) $\downarrow$ & Coverage (\%) $\uparrow$\\
    \midrule
    \textit{Continuous}      & \textbf{0.798} (0.089) & 0.767  & 26.71 \\
    \textit{Discrete}        & 2.048 (0.066) & 1.504   & 24.62 \\
    \textit{Discrete$^+$} (1)& 1.516 (0.077) & 2.031  & 14.82 \\
    \textit{Discrete$^+$} (2)& 1.708 (0.073) & 2.375  & 16.16 \\
    \textit{Discrete$^+$} (3)& 1.856 (0.099) & 2.608  & 16.83 \\
    \textit{Hybrid} (1)      & \underline{0.820} (0.085) & \underline{0.650}  & 25.34 \\
    \textit{Hybrid} (2)      & 0.883 (0.075) & \textbf{0.629}  & \underline{27.30} \\
    \textit{Hybrid} (3)      & 0.935 (0.087) & 0.807  & \textbf{28.73} \\
  \bottomrule
\end{tabular}
}
\end{table}
\end{small}

We measure 800,000 transitions, initializing each 30-second episode randomly from a state sampled from the reference.
For models employing RVQ, we experiment with varying numbers of codebooks as described in Sec.~\ref{sec:04_03_training}.

The results in Table~\ref{tab:prior_rollout} indicate that our \textit{hybrid} representation can generate plausible yet diverse motions from sampling the latent variables.
While it is expected that \textit{continuous} models generate motions that best match the reference motion, our \textit{hybrid} model can synthesize high-quality motion showing comparable trajectory distances.
While RVQ also enhances the performance of \textit{discrete} models, our proposed technique with MQ plays an essential role in generating plausible motions.
Significantly, ours outperforms the \textit{continuous model} regarding the validity and diversity of generated motions.
Our model additionally offers user controllability for balancing the quality and the diversity of motion with RVQ layers, which the \textit{continuous} model cannot provide.
The user can increase the number of codebooks to enhance the diversity while sacrificing minimal imitation quality.
This result also implies extra controllability for task learning. 
For example, if the high-level policy employs a large number of codebooks, the agent can achieve a high level of exploration. 
Conversely, using fewer codebooks allows the agent to behave more safely with smoother trajectories, resulting in more stable training. 
By balancing these properties, we can determine the optimal number of codebooks for each downstream task.

\subsection{Downstream Tasks}\label{sec:downstream_tasks}
\begin{small}
\begin{table*}
\centering
  \caption{Evaluation on the motion in-betweening results. We randomly sample 5 keyframes from a 4 second-length motion sequence, \textit{i.e.}, the average time intervals between keyframes are 1 second. 
  We report two sets of statistics; evaluated only with the keyframes ("Only Keyframes") and with full frames ("Full Frames"). 
  The units for MPJPE and G-MPJPE are $mm$, while units for velocity and acceleartion error are \mbox{mm/frame} and \mbox{mm/frame$^2$}, respectively. We additionally mark the second best score with underlines.}
  \label{tab:mib_eval1}
  \resizebox{0.85\textwidth}{!}{
  \begin{tabular}{c|cccc|cccc}
    \toprule
     & \multicolumn{4}{c|}{Only Keyframes} & \multicolumn{4}{c}{Full Frames} \\
    \toprule
    Models & MPJPE $\downarrow$ & G-MPJPE $\downarrow$ & Vel Error $\downarrow$ & Accel Error $\downarrow$ & MPJPE $\downarrow$ & G-MPJPE $\downarrow$ & Vel Error $\downarrow$ & Accel Error $\downarrow$\\
    \midrule
    \textit{Continuous}      & 157.3 (109.8) & 414.1 (540.3) & 48.44 (19.92) & 30.49 (10.95) & 181.8 (89.74) & 461.6 (502.9) & 43.79 (18.83) & 29.61 (10.87)\\
    \textit{Discrete}        & 111.1 (88.59) & \textbf{304.9} (362.3)  & 36.22 (18.24) & 16.02 (6.771) & 132.5 (79.59) & 348.7 (339.9) & 31.74 (16.57) & 15.69 (6.825)\\
    \textit{Discrete$^+$} & \underline{110.5} (88.11) & 329.5 (346.4)  & \underline{35.02} (18.71) & \underline{13.08} (6.294)& \underline{132.0} (77.24) & \underline{369.8} (312.4) & \underline{30.76} (16.70) & \underline{13.26} (6.311)\\
    \textit{Hybrid}     & \textbf{103.3} (87.98) & \underline{316.7} (392.7)  & \textbf{34.09 (19.97)} & \textbf{12.63 (6.323)} & \textbf{121.1 (79.35)} & \textbf{350.7 (356.6)} & \textbf{29.08 (17.21)} & \textbf{12.46 (6.176)}\\
  \bottomrule
\end{tabular}
}
\end{table*}
\end{small}

Our hybrid model provides versatile latent space that an agent can efficiently navigate and robustly adapt to diverse downstream tasks.
We deliberately demonstrate challenging scenarios where agents must deal with temporally or spatially sparse goal conditions.
We further provide detailed explanations, including the reward functions, in the Appendix.

\subsubsection{Motion In-betweening}
As our first downstream task, we present a motion in-betweening example in which agents must interpolate temporally sparse keyframes throughout the episodes.
During the training, we randomly select zero-velocity poses as keyframes from the original motion sequence. 
Then, we provide the agent with a task goal vector $g$ based on the two keyframes for each local interval.
We enhance the goal observation by augmenting the keyframe states with the time-to-arrival information for the next keyframe~\cite{harvey2020robust,gopinath2022motion}.
We train our agent to mimic the ground-truth trajectories of the reference motion sequence using a full-body imitation reward~\cite{won2020scalable}.

\begin{figure*}[t]
  \centering
    \includegraphics[clip=true, width=\linewidth]{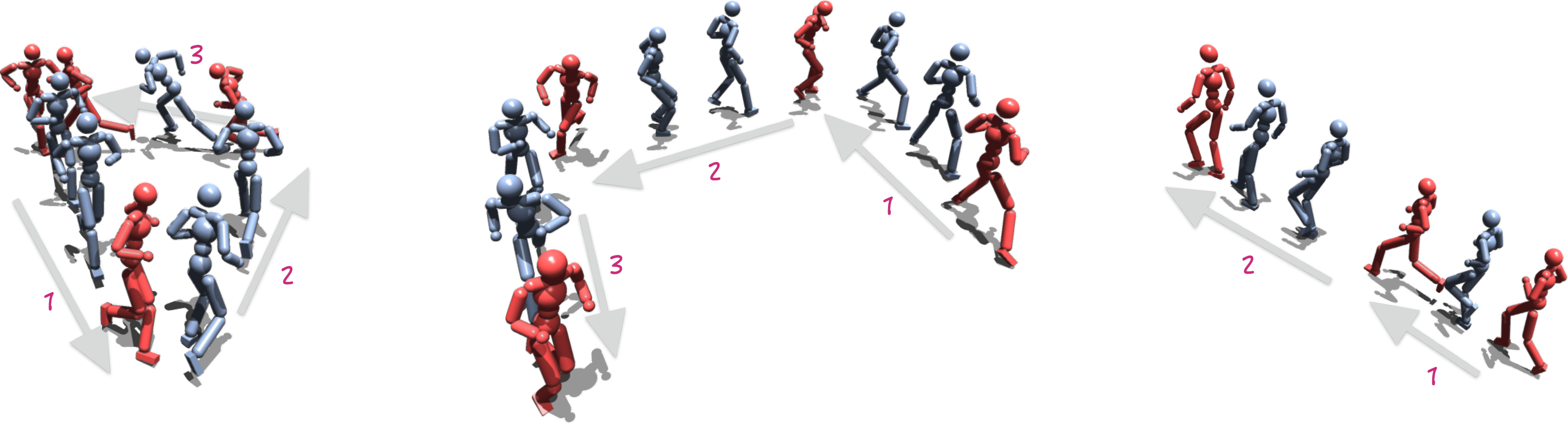}
    \caption{Results of motion in-betweening example. Our simulated character (blue) generates plausible trajectories given zero-velocity keyframes (red). Arrows and numbers indicate the temporal flow of each episode.}
    \label{fig:motion_inbetweening}
\end{figure*}

Our agent successfully interpolates keyframes even in novel scenarios that were not experienced during the training.
To compare our model with the baselines, we quantitatively evaluate the generated motions with the conventional body tracking metrics~\cite{luo2023universal}: mean per joint position error (MPJPE), MPJPE in global coordinates (G-MPJPE), root velocity error, and root acceleration error.
The first two metrics indicate the motion fidelity compared to the ground truth trajectory, whereas the last two metrics represent the smoothness of motion.
Table~\ref{tab:mib_eval1} summarizes the results of 2048 episodes from the test set.
The result demonstrates that our model outperforms all other baselines. 
Notably, the performance of the \textit{continuous} model reveals its failure to generate plausible motions, even with a dense full-body reward calculated from the reference trajectory.

Additionally, we empirically verify that the high-level policy of the \textit{continuous} model is overly sensitive to hyperparameters during training. 
For example, as shown in Figure~\ref{fig:training_curve} (right), the agent engages in excessive exploration when the noise level of the policy is set too high, while it focuses too much on exploitation when the noise level is set too low.
In contrast, models with a discrete prior do not require such modification to balance exploration and exploitation.
This implies that the discrete latent representation significantly enhances training efficiency for the task-learning phase.
Among the models with discrete motion prior, our model generates minimal motion artifacts compared to \textit{discrete} and \textit{discrete$^+$} models, given low values for the velocity and acceleration errors.
We visualize the qualitative results from our model in Figure~\ref{fig:motion_inbetweening} and the supplementary video.
Unlike the baselines, our agent consistently performs natural transitions between keyframes, even when the agent must rapidly shift the direction. 
Furthermore, our agent demonstrates natural motions in challenging scenarios, such as when it is required to walk backward. 

\subsubsection{Head-mounted Device Tracking}
In addition, we explore a challenging task where an agent reconstructs full-body motion given spatially sparse tracking targets.
In this setup, only the head trajectories are provided as a conditioning signal for the agents.
The goal observation is provided as the state difference between the target and simulated head positions.
The agent additionally observes future head trajectories for four additional time steps to resolve inherent ambiguity of the extremely sparse observation.
During training, we calculate the reward based on how closely the agent's head state follows the target trajectory.

\begin{small}
\begin{table}
\centering
  \caption{Evaluation on the head-mounted device tracking. Similar to the data presented in Table~\ref{tab:mib_eval1}, we utilize body tracking metrics to quantitatively evaluate the quality of motion.}
  \label{tab:track_eval}
  \resizebox{\linewidth}{!}{
  \begin{tabular}{c|cccc}
    \toprule
    Models & MPJPE $\downarrow$ & G-MPJPE $\downarrow$ & Vel Error $\downarrow$ & Accel Error $\downarrow$\\
    \midrule
    \textit{Continuous}      & 168.7 (43.64) & 174.3 (108.5) & 29.10 (15.03) & 23.53 (11.22)\\
    \textit{Discrete}        & \underline{100.4} (47.91) & \textbf{110.9} (107.1)  &  22.40 (10.55) & 15.34 (6.102)\\
    \textit{Discrete$^+$} & 103.7 (50.39) & 116.1 (144.7)  & 21.42 (10.26) & \underline{14.19} (5.700)\\
    \textit{Hybrid}     & \textbf{96.90} (57.88) & \underline{112.9} (141.6)  & \textbf{20.46} (11.57) & \textbf{13.48} (5.930)\\
  \bottomrule
\end{tabular}
}
\end{table}
\end{small}
\begin{figure*}[t]
  \centering
    \includegraphics[clip=true, width=\textwidth]{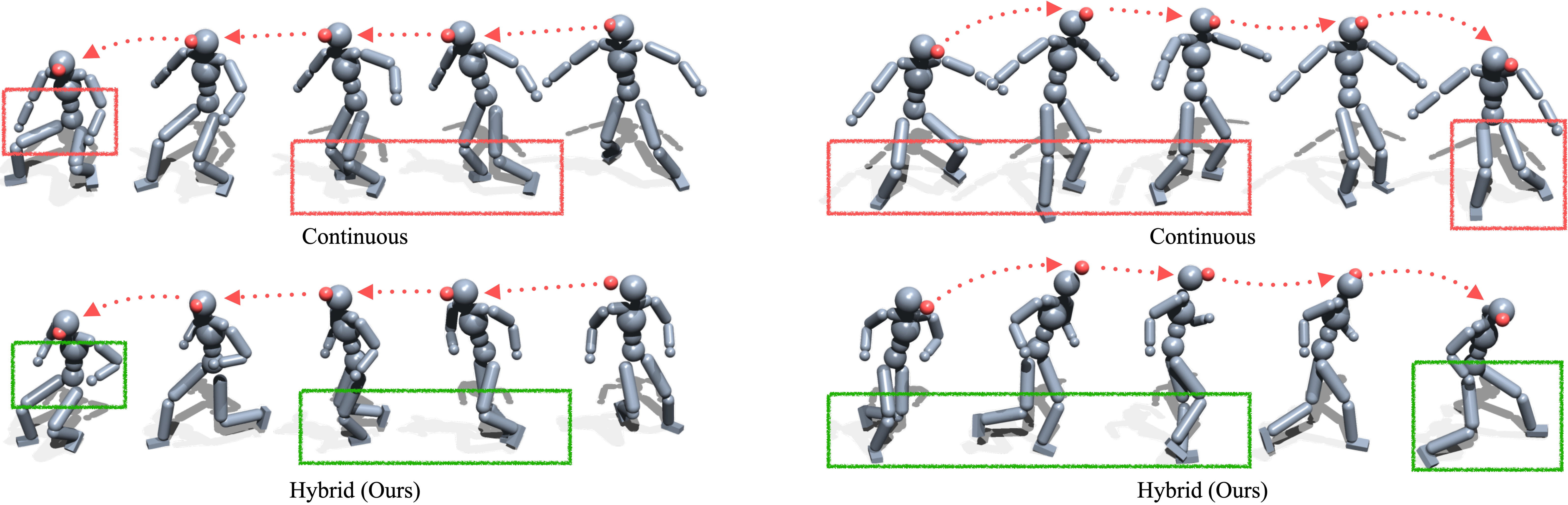}
    \caption{Qualitative comparison between \textit{continuous} and \textit{hybrid} models on head-mounted device tracking example. Our agent faithfully adheres to the learned motion prior while \textit{continuous} model cannot. }
    \label{fig:track_viz1}
\end{figure*}

\begin{figure*}[t]
  \centering
    \includegraphics[clip=true, width=\textwidth]{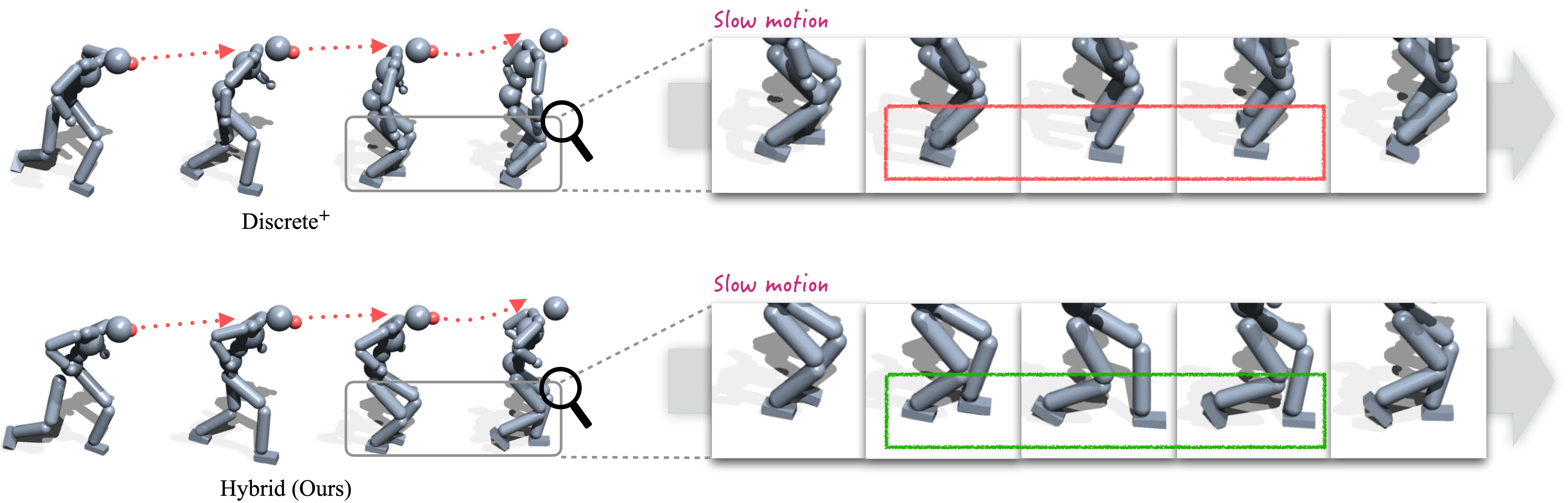}
    \caption{Qualitative comparison between \textit{discrete$^+$} and \textit{hybrid} models on head-mounted device tracking example. While \textit{discrete$^+$} model sometimes violates natural gaits, our model consistently exhibits realistic locomotion.}
    \label{fig:track_viz2}
\end{figure*}
Similar to the temporally sparse setting, we find that our agent successfully discovers the optimal set of motions for head-mounted device tracking. 
Quantitative results in Table~\ref{tab:track_eval} and qualitative results in Figure~\ref{fig:track_viz1} consistently demonstrate that our model effectively reuses the pretrained motion prior. 
The \textit{hybrid} latent space enables the agent to track the head trackers more accurately than the \textit{continuous} latent space, indicating enhanced training efficiency.
Surprisingly, the gap between ours and the \textit{discrete} models becomes more apparent in this scenario compared to the motion in-betweening.
We attribute this to the robustness of our model under environments with sparse rewards.
As shown in Figure~\ref{fig:track_viz2}, our model maintains a natural gait pattern, while \textit{discrete$^+$} model fails to do so.
As clearly demonstrated with the supplementary video, \textit{discrete} models often experience highly jittery movements as exploring the quantized latent space to fulfill the sparse goal.
Such movement significantly degrades the visual quality of the generated motion. 
In contrast, our model does not suffer from these artifacts and consistently adheres to the learned motion priors.

\subsubsection{Point-goal Navigation}
Lastly, we demonstrate a point-goal navigation example, which demonstrates a setting with spatio-temporally sparse goals. 
We provide the relative position to the goal in the ground plane as a goal observation.
We employ \textit{target location} reward that measures the relative distance and the heading velocity, similarly with the previous approach~\cite{peng2018deepmimic}.
Once the agent arrives at the goal, we give the maximum value for the reward to prevent meaningless motions in the goal location.


\begin{figure*}[t]
  \centering
    \includegraphics[clip=true, width=\textwidth]{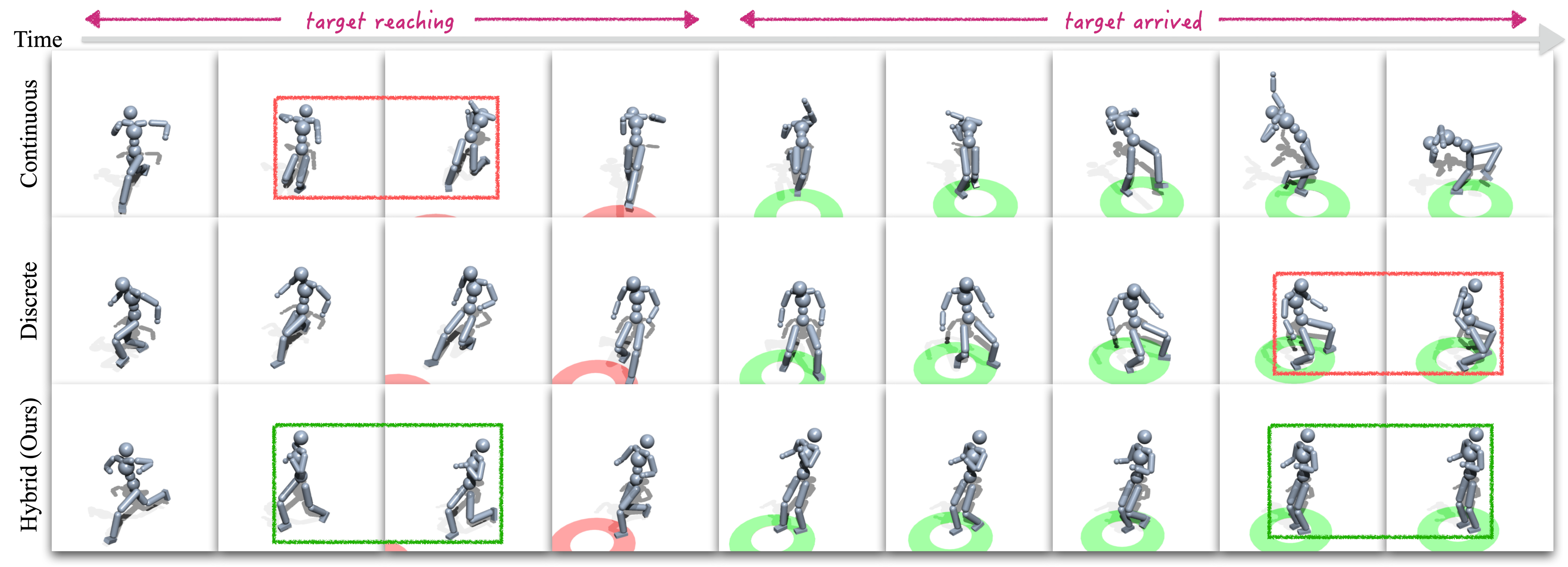}
    \caption{Qualitative comparison between baselines (\textit{continuous}, \textit{discrete}) and \textit{hybrid} model on the point-goal navigation example. \textit{Continuous} model prioritizes speed over motion quality to maximize the task reward, which is higher when the goal is reached quickly.
    On the other hand, \textit{discrete} model produces plausible motion when reaching the target location but struggles to maintain natural idle motion once it arrives.}
    \label{fig:nav_viz}
\end{figure*}
\begin{figure}[t]
    \centering
    \includegraphics[clip=true, width=0.99\linewidth]{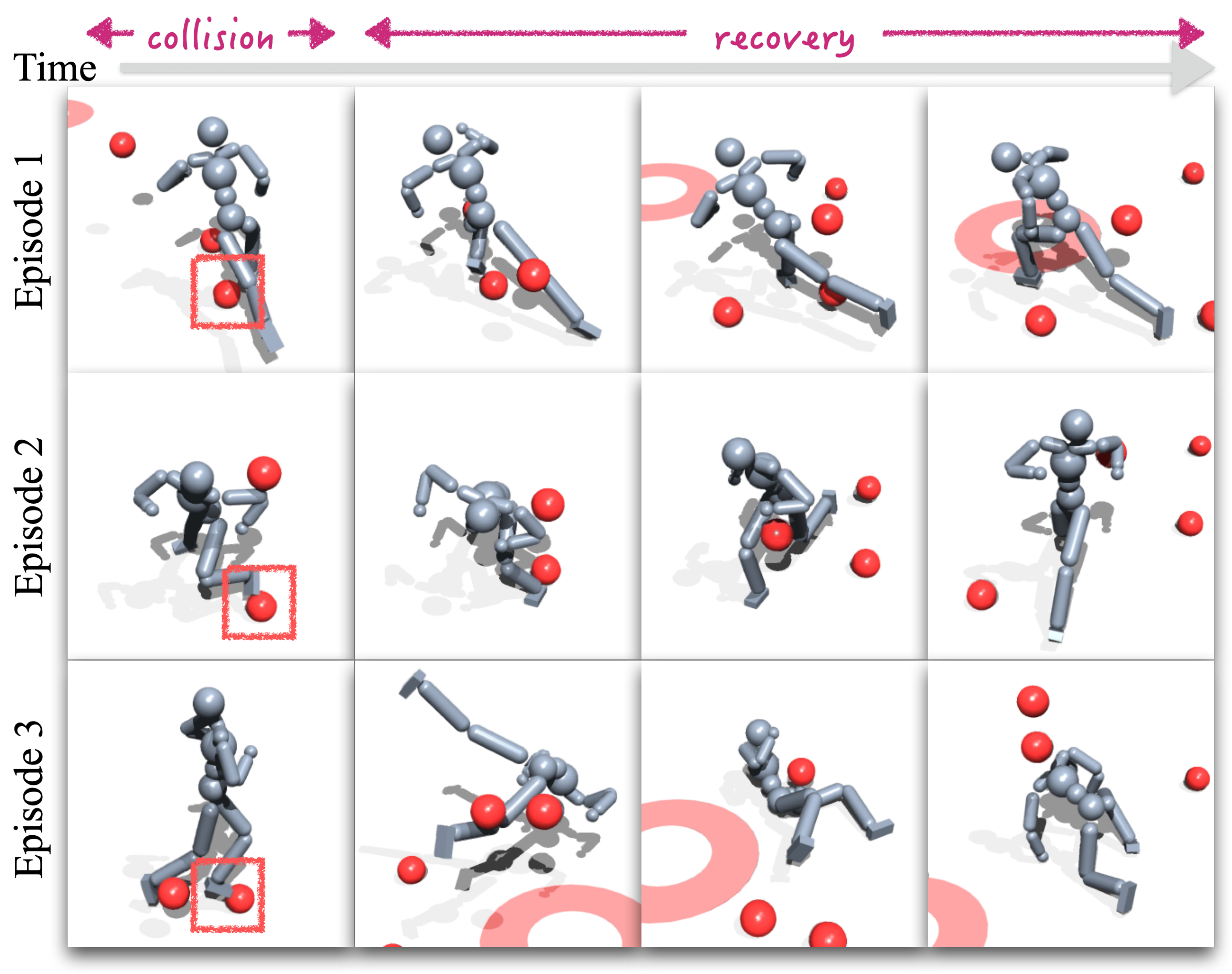}
    \caption{Zero-shot adaptation to unexpected perturbations in a point-goal navigation task. Every 0.3 seconds, obstacles (red spheres) are thrown at the agent from random directions and velocities. Red boxes highlight the key moment of the collision that triggers the largest perturbation in an episode. Our agent consistently maintains natural movement despite significant disruptions.}\label{fig:point_goal_obstacles}
\end{figure}
In this scenario, we can observe a similar tendency to the results from the other tasks.
Note that this example does not include any ground-truth motions for additional guidance, and therefore the agent motions contain the pure exploration results to fulfill the task.
The agents receive the maximum reward as long as they reach the goal, and there is no means to penalize unnatural motions.
As demonstrated in Figure~\ref{fig:nav_viz} and the accompanying video, only our \textit{hybrid} model exhibit natural idle motions, whereas other models generate awkward movements.
The result shows that our model can faithfully generate natural motions with a versatile set of motion prior.
Other works require additional reward designs to encourage desirable idle states, which involve non-trivial hand-crafted terms in the reward function.

We also test the robustness of our policy against physical perturbations. 
We introduce unexpected disruptions in the navigation environment and assess the agent's ability to fulfill the goal stably.
To evaluate such ability, we test the policy trained on plain point-goal navigation environments in a zero-shot manner.
For the external perturbations, spherical objects with a radius of 10 cm and a mass of 4.2 kg are thrown toward the agent every 0.3 seconds.
The objects are thrown from a random location 1 meter away from the characters, with velocities randomly selected between 10 and 20 m/s.
In Figure~\ref{fig:point_goal_obstacles}, we visualize the three different episodes of agents navigating with random perturbations.
We find that our agent successfully maintains balance after collisions. 
Even when knocked down by severe impacts, as illustrated in the third row of the figure, our agent recovers without deviating from the learned motion distribution. 

These results demonstrate that our hybrid latent representation effectively handles test-time variations. 
We attribute this adaptability to the residual design, where the high-level policy modifies only a marginal portion of the latent vector produced by the prior network. 
This design enhances the agent’s ability to respond to physical perturbations with greater stability.
An ablation study from previous work~\cite{luo2023universal} also supports that such residual modeling enhances performance at the task learning phase.
Additional evidence is provided by the unconditional motion generation discussed in Sec.~\ref{sec:06_02_unconditional}. 
As shown in Figure~\ref{fig:prior_sampling}, our \textit{hybrid} model uses a continuous latent vector as an offset to the discrete latent vector, which produces much more stable motions compared to \textit{discrete} models.

\section{Conclusion}
In this paper, we present a novel architecture that utilizes integrated latent representation for the reusable motion prior. 
We suggest a hybrid representation that discretizes only the difference between the posterior and prior latent vectors. 
This approach fully utilizes the expressive yet temporally coherent continuous latent vector from the prior network during the task learning phase.
In addition, we employ Residual Vector Quantization to improve the code usage and the training efficiency with the categorical latent prior.
In the experiments, we demonstrate that our model is versatile in various scenarios.
We show that \textit{hybrid} latent space for motion prior enables efficient training in the task learning phase, allowing agents to exhibit temporally smooth motions consistently.

However, our models did not fully exploit the possible technical enhancement for recent variations of quantization or training pipelines.
We believe state-of-the-art techniques such as finite scalar quantization~\cite{mentzer2023finite} or group RVQ~\cite{yang2023hifi} can alleviate inherent limitations of vector quantization, such as overfitting problem~\cite{wu2020vector}.
Additionally, extending the hybrid latent model to alternative training pipelines, such as model-based approaches~\cite{yao2022controlvae, yao2023moconvq}, could enable a more comprehensive comparison with other state-of-the-art methods, providing valuable insights.

We suggest a few future directions to extend this work.
Extending our hybrid latent model to the multi-subgoal scenarios could be one interesting research direction. 
For instance, future work could adapt the original PPO algorithm to incorporate a multi-task training scheme, as demonstrated in prior work~\cite{xu2023composite}.
Moreover, our model can be helpful in the motion composition task.
By splitting the discrete latent space part-wise, one can achieve a dynamic combination of the part skills while maintaining the whole-body consistency with the continuous latent vector from the prior network.

\section*{Acknowledgements}
This work was supported by the National Research Foundation of Korea(NRF) grant funded by the Korea government(MSIT) (No. RS-2023-00218601), Creative-Pioneering Researchers Program through Seoul National University, INMC, and BK21 FOUR program of the Education and Research Program for Future ICT Pioneers, Seoul National University in 2025.
Jungdam Won is partly supported by the NRF grant(RS-2024-00450647) and the Institute of Information \& Communications Technology Planning \& Evaluation(IITP)-ITRC(Information Technology Research Center) grant (IITP-2025-RS-2020-II201460) funded by the Korea government(MIST).

\section*{Appendix}
\setcounter{section}{0}
\renewcommand\thesection{\Alph{section}}
\section{Ablation Study}
In this section, we provide an ablation study on our \textit{hybrid} model. 

\subsection{Effectiveness of Residual Vector Quantization}
We first demonstrate the effectiveness of Residual Vector Quantization (RVQ) from the perspective of code usage and training efficiency.
For comparison, we additionally train an ablated model where we use a conventional vector quantization layer instead of RVQ, namely \textit{hybrid-vq}.

\begin{figure}[h]
  \centering
  \includegraphics[width=\linewidth]{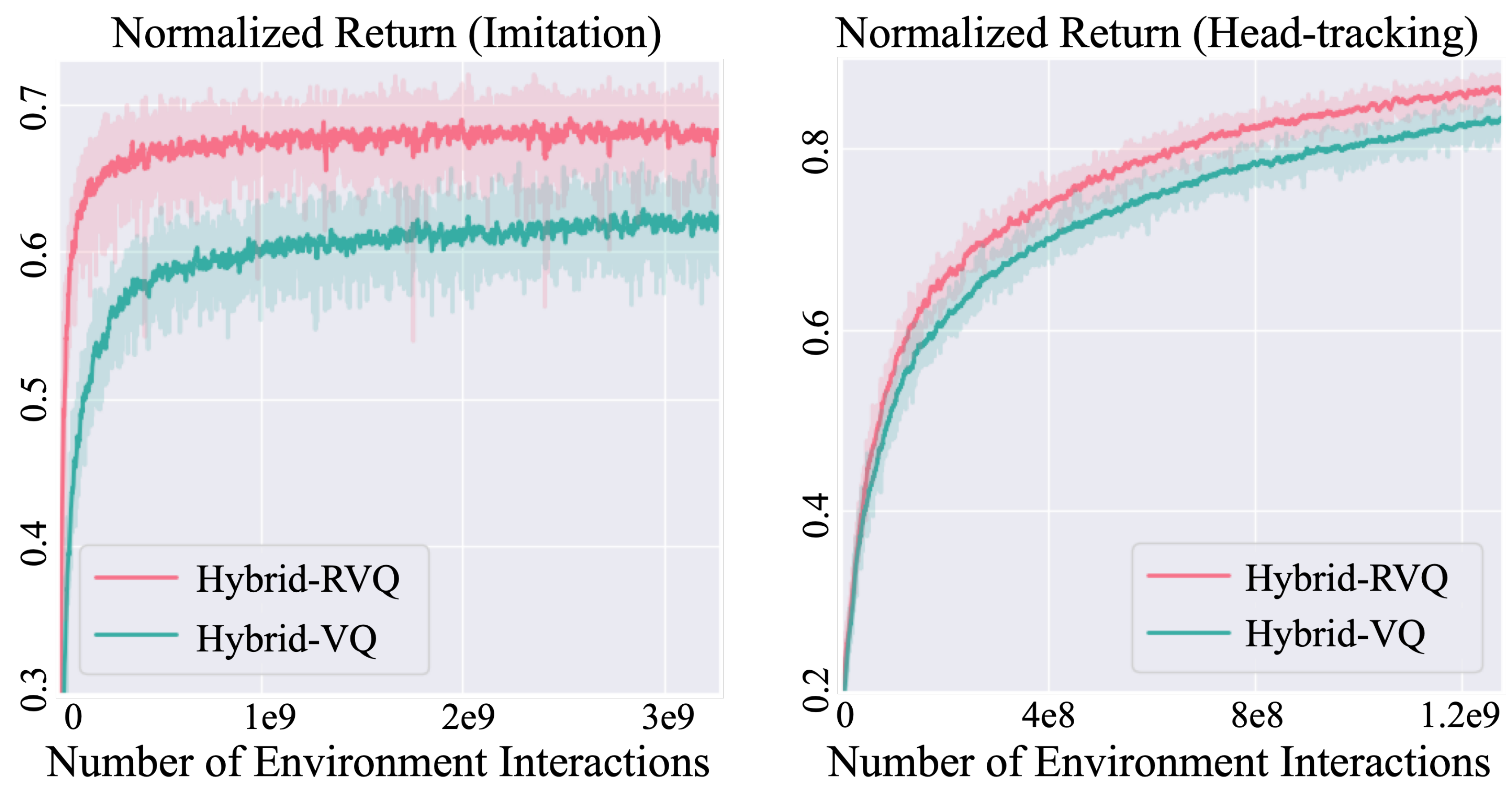}
  \caption{Training curves for imitation learning (\textit{left}) and task learning (head-mounted device tracking, \textit{right}). We allocate 8192 codes for the \textit{hybrid} model with a single VQ layer (Hybrid-VQ), while setting RVQ with 8 codebooks each containing 1024 codes for our main model (Hybrid-RVQ).}
  \label{fig:ablation_vq}
\end{figure}

As shown in the training curve in Figure~\ref{fig:ablation_vq}, \textit{hybrid-vq} model fails to achieve a sufficient level of imitation.
This implies that RVQ contributes to the enhanced code usage for the latent representation given a finite number of discrete codes.

Moreover, during task learning, we find that the \textit{hybrid-vq} model exhibits lower performance in terms of normalized return compared to our model. 
Additionally, the \textit{hybrid-vq} model requires more training time due to the computational burden of the $\text{argmin}$ operation (Eq. (3)) with a larger discrete action space.
In contrast, our model with RVQ handles this more efficiently. 

\subsection{Number of Codebooks for Task Learning}
In unconditional motion generation, we show that our \textit{hybrid} model can balance the quality and the diversity of motion by controlling the number of codebooks used for prior sampling. 
We provide an additional evaluation to clearly demonstrate this balance in Table~\ref{tab:prior_rollout_additional}.

\begin{small}
\begin{table}[h]
\centering
  \caption{Evaluation on the unconditional motion generation. The value inside the parenthesis indicates number of the used codebooks. We additionally mark the second best value with underlines.}
  \label{tab:prior_rollout_additional}
  \resizebox{0.95\linewidth}{!}{
  \begin{tabular}{c|ccc}
    \toprule
    Models & Distance ($d$) $\downarrow$ & Filtered (\%) $\downarrow$ & Coverage (\%) $\uparrow$\\
    \midrule
    \textit{Hybrid} (1)      & \textbf{0.820} (0.085) & \underline{0.650}  & 25.34 \\
    \textit{Hybrid} (2)      & \underline{0.883} (0.075) & \textbf{0.629}  & 27.30 \\
    \textit{Hybrid} (3)      & 0.935 (0.087) & 0.807  & 28.73 \\
    \textit{Hybrid} (4)      & 0.963 (0.090) & 0.738  & 29.29 \\
    \textit{Hybrid} (5)      & 0.980 (0.092) & 0.774  & 29.94 \\
    \textit{Hybrid} (6)      & 0.975 (0.081) & 0.735  & 30.30 \\
    \textit{Hybrid} (7)      & 1.012 (0.083) & 0.772  & \underline{30.83} \\
    \textit{Hybrid} (8)      & 1.017 (0.089) & 0.749  & \textbf{30.99} \\
  \bottomrule
\end{tabular}
}
\end{table}
\end{small}

\begin{figure}[h]
  \centering
  \includegraphics[width=\linewidth]{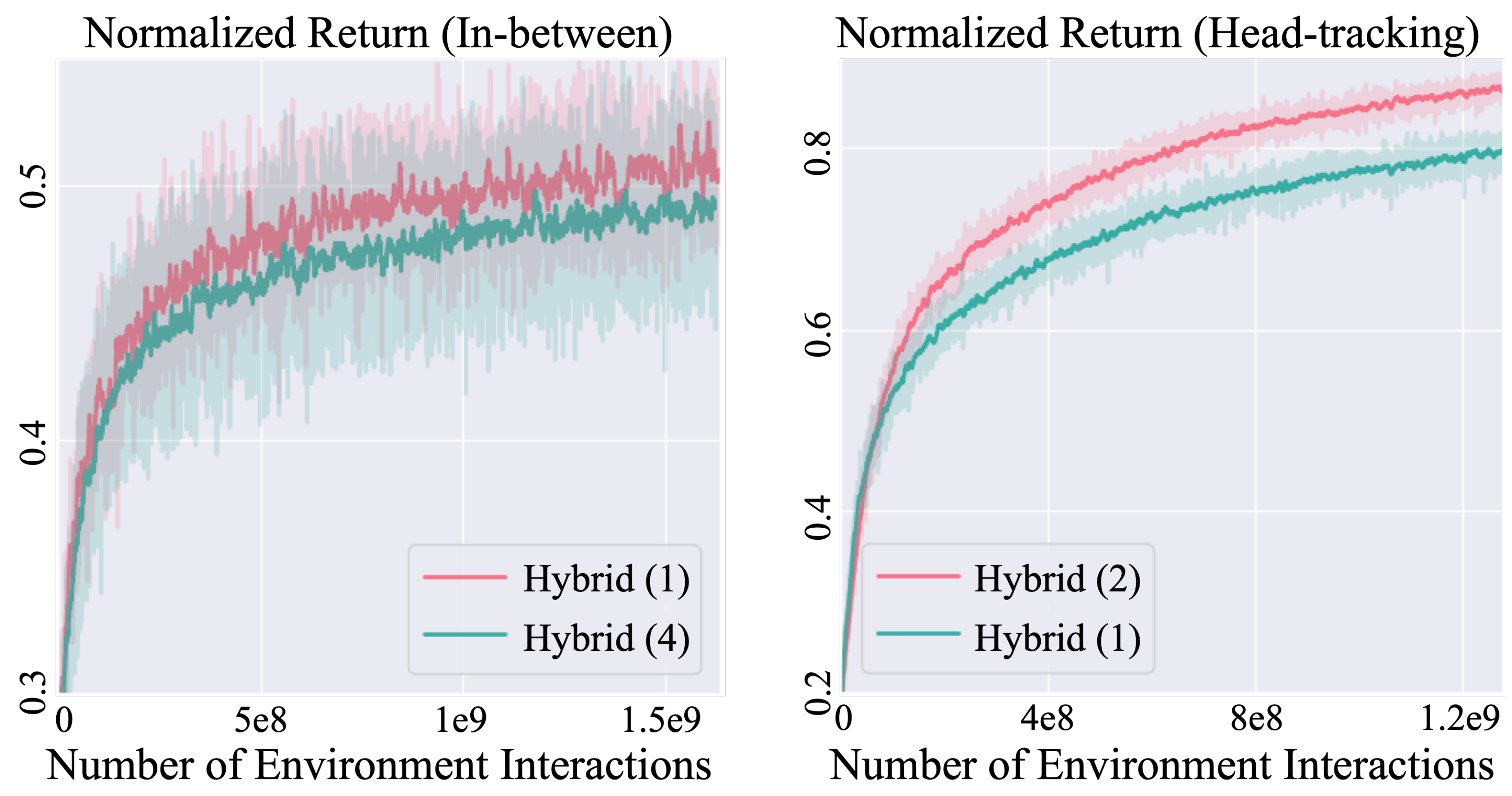}
  \caption{Training curves for motion in-betweening (\textit{left}) and head-mounted device tracking (\textit{right}). We indicate a number of active codebooks used for training in the parenthesis.}
  \label{fig:ablation_codebook}
\end{figure}

Along with this result, RVQ offers additional controllability through the number of active codebooks during the task learning phase.
We show the result from different numbers of codebooks employed for task learning in Figure~\ref{fig:ablation_codebook}.
As demonstrated in the results for motion in-betweening, using fewer codebooks leads to less diverse but more precise motion, allowing the high-level policy to be trained more stably. 
Conversely, in the tracking example, utilizing more codebooks can sometimes yield better results by enabling the high-level policy to explore the action space more actively during training. 
However, employing too many codebooks may excessively enlarge the action space, increasing the training complexity for task learning. 
Balancing this trade-off, we select the optimal number of active codebooks for each downstream task, achieving the best performance among the baselines.

\section{Imitation Learning Details}
In this section, we provide an additional description of the imitation learning phase.
We first elaborate on the further details about the imitation learning phase (Sec.~\ref{sec:a1_training}).
Then, we provide additional information on the expert policy imported for training our imitation policy (Sec.~\ref{sec:a2_expert}).

\subsection{Training}\label{sec:a1_training}
In Figure 2, our imitation policy includes a posterior network that outputs a latent vector $ z $ given the current simulation state $ s $ and the target state $\tilde{s} $ sampled from the dataset. 
As represented in Eq.~(7), the current simulation state $ s $ primarily encodes the agent's proprioceptive state. 
In contrast, the target state $\tilde{s} $ represents the tracking objective.
Instead of directly using the target state vector from the dataset, we represent $\tilde{s} $ as the state difference between the current state and the target state, following the approach in previous work \cite{luo2023perpetual, luo2023universal}. 
Specifically, we first calculate the kinematic states for the $ J $ rigid bodies of the agent, which include positions $ p \in \mathbb{R}^{J \times 3} $, velocities $v \in \mathbb{R}^{J \times 3} $, orientations $\theta \in \mathbb{R}^{J \times 4} $, and angular velocities $\omega \in \mathbb{R}^{J \times 3} $.
Simultaneously, we extract the same configurations from the target state of the reference motions, which we represent as $[\tilde{p}, \tilde{v}, \tilde{\theta}, \tilde{\omega}]$.
Then, we calculate the state difference as follows:
\begin{equation}
    \tilde{s}=[p - \tilde{p}, v - \tilde{v}, \theta \ominus \tilde{\theta}, \omega - \tilde{\omega}],
    \label{eq:state_diff}
\end{equation}
where $\ominus$ denotes the operation to measure the difference between quaternions.

To improve sample efficiency, we also employ the early termination technique~\cite{peng2018deepmimic}, which is to terminate an episode when an agent violates a certain condition.
For our imitation learning, we set an early termination condition that properly prevents large deviations from the ground-truth trajectories. 
Following a similar approach as in previous work \cite{luo2023perpetual,luo2023universal}, we set a termination condition based on the body positions. 
Specifically, we terminate an episode when the maximum distance from each body part of the agent to the corresponding body part in the target pose exceeds 0.5 meters, \textit{i.e.,} ensuring that all body positions remain within 0.5 meters of the ground truth.

\subsection{Expert Policy}\label{sec:a2_expert}
Following the previous work~\cite{luo2023universal}, we employ \textit{online distillation} for training the imitation policy (Sec. 4.3).
Unlike conventional Reinforcement Learning (RL), online distillation trains policy with supervised learning, resulting in much faster training compared to RL.
To utilize online distillation, we need to train an expert policy in advance.

Basically, any method that achieves a reliable level of imitation can be adopted for training the expert policy. 
For example, one can employ an ensemble policy as an expert policy following previous works~\cite{luo2023perpetual, luo2023universal} that have achieved 100\% imitation for extremely large datasets like AMASS~\cite{mahmood2019amass}.
However, we empirically find that our target dataset, LaFAN1 \cite{harvey2020robust}, does not require such a complex process for training the expert policy. 
Instead, we choose a simpler encoder-decoder structure (Section 3.1) to serve as the architecture for the expert policy.
Due to the absence of the regularization on the latent space, this simple architecture cannot learn an effective latent space that is optimal for the high-level policy.  
However, we find that this type of architecture supports stable training of policy since such model does not require to optimize additional loss terms such as \textit{Kullback-Leibler} loss in Eq.~(1) or \textit{commitment} loss in Eq.~(3).

To train an expert policy, we leverage an imitation reward function from the previous work~\cite{won2020scalable}.
Specifically, our imitation reward is multiplication of four terms, where each term calculates tracking error for root position $p_ {\text{root}}$, joint position $q$, joint velocity $\alpha$, and the positions of end-effectors $p_{\text{ee}}$, respectively.
The total imitation reward term $r_{\text{im}}$ can be written as
\begin{equation}
    \begin{gathered}
    r_{\text{im}} = r_{\text{root}}\cdot r_{q} \cdot r_{\alpha} \cdot r_{\text{ee}},\\
    r_{root} = \exp({-10 \cdot \|p_{\text{root}}-\tilde{p}_{\text{root}}\|^2}),\\
    r_{q} = \exp({-2 \cdot \|q - \tilde{q}\|^2}),\\
    r_{\alpha} = \exp({-0.5 \cdot \|\alpha - \tilde{\alpha}\|^2}),\\
    r_{\text{ee}} = \exp({-40 \cdot \|p_{\text{ee}} - \tilde{p}_{\text{ee}}\|^2}),
    \end{gathered}\label{eq:imitation_reward}
\end{equation}
where symbols with the \textit{tilde} indicate value from the reference state.

In addition to the imitation reward, we incorporate style reward from the adversarial training pipeline~\cite{peng2021amp, luo2023perpetual}.
In this setting, a trainable discriminator $D_\phi$ determines whether the queried state transition $(s, s')$ comes from the simulated agent or the reference dataset. 
The discriminator outputs a value close to 1 if it predicts that the transition likely originates from the agent, and a value close to 0 for the opposite case.
We express style reward $r_{\text{style}}$ and the total reward $r$ as
\begin{equation}
\begin{gathered}
      r_{\text{style}}=-5\cdot\log{(1-D_\phi(s,s'))},\\
      r = 0.5 \cdot r_{\text{im}} + 0.5 \cdot r_{\text{style}}.
\end{gathered}
\end{equation}\label{eq:amp_style_reward}
\section{Task Learning Details}
Expanding Section 6.3, we describe detailed settings for the downstream tasks. 
The explanation for each task includes mathematical expressions of task-specific goal observation, reward function, and the condition for early termination of the training.

\subsection{Motion In-betweening}
In this example, an agent learns to generate interpolated motions given two keyframes and the queried time interval as goal observation $g$.
In each episode, we randomly select 4 keyframes from a 4-second motion sequence sampled from the dataset.

\paragraph*{Goal observation}
We employ the configuration of the goal observation vector used in the previous work~\cite{gopinath2022motion}.
An agent observes two consecutive keyframes where one is the start keyframe $f_s$ and the other is the goal keyframe $f_g$ for each interval.
Given an agent with $ J $ rigid bodies, each keyframe $f$ contains body position $ p_f \in \mathbb{R}^{J \times 3} $ and orientation $\theta_f \in \mathbb{R}^{J \times 4} $ in the global coordinate. 
At each timestep, the goal observation is calculated by measuring the difference between each keyframe and the simulated state, represented as
\begin{equation}
\begin{gathered}
    \bar{p}_f = p - p_f, \\
    \bar{\theta}_f = \theta \ominus \theta_f,
\end{gathered}
\end{equation}

where $ p $ and $ \theta $ denote the body position and orientation of the simulated agent.
We repeat this process for the start keyframe $f_s$ and the goal keyframe $f_g$, which in result obtains a vector of $\hat{g} = [\bar{p}_{f_s}, \bar{\theta}_{f_s}, \bar{p}_{f_g}, \bar{\theta}_{f_g}]$.
Followed by the previous work~\cite{harvey2020robust}, we then augment the goal observation with a positional encoding of time-to-interval $e_{\text{time}}$, which we write as
\begin{equation}
    g = \hat{g} + e_{\text{time}}.
\end{equation}

\paragraph*{Reward}
To encourage reliable motion, we guide the agent with the full-body imitation reward.
We utilize the similar imitation reward term in Eq.~(\ref{eq:imitation_reward}).
However, since the goal observation is sparse, it is impossible for the agent to strictly follow the ground-truth trajectories. 
Therefore, we modify the reward from a multiplication format to a summation format in order to encourage agents to behave flexibly.
We represent the modified reward as 
\begin{equation}
    r = 0.1\cdot r_{\text{root}} + 0.65 \cdot r_q + 0.1 \cdot r_\omega + 0.15 \cdot r_{\text{ee}}.
\end{equation}

\paragraph*{Early Termination}
To prevent an agent from interpolating in an unrealistic manner, we set an early termination condition that properly prevents large deviations from the ground-truth trajectories. 
Therefore, we employ the same condition used in the imitation learning phase.

\subsection{Head-mounted Device Tracking}
In this scenario, we train an agent to reconstruct the full-body motion given the target head trajectories.
For each episode, we sample 5 seconds of target head trajectories from the reference motion clips.

\paragraph*{Goal Observation}
We can treat imitation learning as full-body tracking since it basically encourages agents to follow the target trajectories.
In contrast to full-body tracking, this scenario assumes that only the target head trajectories are available for the agent.
Therefore, we construct the goal observation vector $g$ by selecting the head parts from the target reference state $\tilde{s}$ in Eq.~(9).

\paragraph*{Reward}
We employ a similar reward structure used for motion in-betweening. 
However, we modify the total reward term to consider only the head part, allowing other parts to move freely without being penalized. 
Even though the reward term does not regulate the style of full-body motion, our model can still generate natural full-body motions.

\paragraph*{Early Termination}
Similarly, with the goal observation and the reward, we only account for the head deviations in the early termination condition.
We terminate an episode when the head of the simulated agent deviates from the ground truth head position more than 1m.

\subsection{Point-goal Navigation}
In this scenario, the agent is trained to reach a target location that is randomly respawned within 10 meters from the agent's initial position.
Additionally, we set the target location to be updated every 4 seconds. 
Since the maximum episode length is 8 seconds, the agent can experience at most one target change per training episode.

\paragraph*{Goal Observation}
Since the target location is on the ground plane, we allow the agent to know its relative position toward the goal.
The goal observation can be abbreviated as
\begin{equation}
    g = p_{xy} - t_{xy}, 
\end{equation}
where $p_{xy}, t_{xy}\in \mathbb{R}^2$ indicate the locations of the agent and the target on the ground, respectively.

\paragraph*{Reward}
We employ the reward function tailed for the \textit{target location} task in the previous work~\cite{peng2018deepmimic}.
The reward consists of two parts: the first term $r_{\text{pos}}$ measuring positional error from the target location, and the second term $r_{\text{speed}}$ measuring the heading speed.
Particularly, $r_{\text{speed}}$ encourages the agent to move toward the target location but does not provide an extra reward once the agent achieves the minimum target speed.
Given the minimum target speed of $u = 1~(m/s)$, we can represent the total reward function as
\begin{equation}
    \begin{gathered}
    r = 0.7 \cdot r_{\text{pos}} + 0.3 \cdot r_{\text{speed}},\\
    r_{\text{pos}} = \exp[-0.5 \cdot \|p_{xy} - t_{xy}\|^2],\\
    r_{\text{speed}} = \exp[-\max(0, u - \langle v_{xy}, {{p_{xy} - t_{xy}}\over{\|p_{xy} - t_{xy}\|}}\rangle)^2],
    \end{gathered}
\end{equation}
where $v_{xy} \in \mathbb{R}^2$ denotes the current velocity of the agent on the ground.

\paragraph*{Early Termination}
Unlike the previous tasks, this scenario does not include any reference trajectories sampled from the dataset.
Therefore, we simply terminate the episode when the agent's head position falls below 0.5 meters, which potentially indicates a fall state.
\section{Hyperparameters}\label{sec:hyperparameters}
We provide hyperparameters for training.
Note "$\rightarrow$" indicates scheduling in Table~\ref{tab:hyp_imi}-\ref{tab:hyp_nav}.
We also mark the parameters that are only applicable to certain model with the parenthesis, \textit{e.g.,} $\mathcal{L}_{\text{KL}}$ for \textit{continuous} model.
\begin{table}[t]
  \caption{Hyperparamters used for imitation learning.}
  \label{tab:hyp_imi}
  \resizebox{0.9\linewidth}{!}{
  \begin{tabular}{c|c}
    \toprule
    Parameters & value \\
    \midrule
    horizon length & 16\\
    number of environment   & 8192\\
    batch size & 65536 \\
    learning rate   & 2e-4\\
    weight for $\mathcal{L}_{\text{expert}}$ & 10\\
    weight for $\mathcal{L}_{\text{KL}}$ (\textit{continuous})& 0.05 $\rightarrow$ 0.5 \\
    weight for $\mathcal{L}_{\text{mm}}$ (\textit{hybrid})& 0.1 $\rightarrow$ 1.0 \\
    weight for $\mathcal{L}_{\text{commit}}$ (\textit{discrete}, \textit{hybrid}) & 1.0 \\
    weight for $\mathcal{L}_{\text{reg}}$ (\textit{continuous}) & 0.005 \\
    weight for $\mathcal{L}_{\text{reg}}$ (\textit{discrete}, \textit{hybrid}) & 0.05 \\
    codebook size (\textit{discrete}) & 8192\\
    codebook size (\textit{discrete$^+$}, \textit{hybrid}) & 1024\\
    Number of codebooks (\textit{discrete$^+$}, \textit{hybrid}) & 8\\
  \bottomrule
\end{tabular}
}
\end{table}

\begin{table}[h]
  \caption{Hyperparamters used for motion in-betweening.}
  \label{tab:hyp_mib}
  \resizebox{0.9\linewidth}{!}{
  \begin{tabular}{c|c}
    \toprule
    Parameters & value \\
    \midrule
    horizon length & 16\\
    number of environment  & 4096\\
    batch size & 32768 \\
    learning rate   & 5e-5\\
    $\gamma$ for PPO & 0.99\\
    $\tau$ for PPO & 0.95\\
    clip range for PPO & 0.1\\
    Number of active codebooks (\textit{discrete$^+$}, \textit{hybrid}) & 1\\
  \bottomrule
\end{tabular}
}
\end{table}

\begin{table}[h]
  \caption{Hyperparamters used for head-mounted device tracking.}
  
  \label{tab:hyp_tracking}
  \resizebox{0.9\linewidth}{!}{
  \begin{tabular}{c|c}
    \toprule
    Parameters & value \\
    \midrule
    horizon length & 16\\
    number of environment  & 8192\\
    batch size & 65536 \\
    learning rate   & 5e-5\\
    $\gamma$ for PPO & 0.99\\
    $\tau$ for PPO & 0.95\\
    clip range for PPO & 0.1\\
    Number of active codebooks (\textit{discrete$^+$}, \textit{hybrid}) & 2\\
  \bottomrule
\end{tabular}
}
\end{table}

\begin{table}[h]
  \caption{Hyperparamters used for point-goal navigation.}
  \label{tab:hyp_nav}
  \resizebox{0.9\linewidth}{!}{
  \begin{tabular}{c|c}
    \toprule
    Parameters & value \\
    \midrule
    horizon length & 16\\
    number of environment  & 8192\\
    batch size & 65536 \\
    learning rate   & 5e-5\\
    $\gamma$ for PPO & 0.99\\
    $\tau$ for PPO & 0.95\\
    clip range for PPO & 0.1\\
    Number of active codebooks (\textit{discrete$^+$}, \textit{hybrid}) & 1\\
  \bottomrule
\end{tabular}
}
\end{table}
\bibliographystyle{eg-alpha-doi} 
\bibliography{egbibsample}       




\end{document}